\documentclass{article}
\usepackage{style}
\usepackage{caption}
\usepackage[affil-it]{authblk}

\title{\textbf{Integrating Randomized Controlled Trial and External Control Data Using Balancing Weights: A Comparison of Estimands and Estimators}}

\author[1]{Peijin Wang}

\author[1,2]{Hwanhee Hong}

\author[3]{Kyungeun Jeon}

\author[1,2]{Laine Elliott Thomas\thanks{corresponding author: Laine Elliott Thomas, 200 Morris Street, Durham, NC 27701, USA. 
Email: laine.thomas@duke.edu}}

\affil[1]{Department of Biostatistics and Bioinformatics Duke University, NC, USA}

\affil[2]{Duke Clinical Research Institute, NC, USA}

\affil[3]{Department of Biostatistics, Johns Hopkins University, MD, USA}

\newtheorem{assumption}{Assumption}

\providecommand{\keywords}[1]{\textbf{\textit{Keywords---}} #1}

\begin{document}
\maketitle

\begin{abstract}
Randomized controlled trials (RCTs) face inherent limitations, such as ethical or resource constraints, which lead to a limited number of study participants. To address these limitations, recent research endeavors have sought to incorporate external control (EC) data, such as historical trial data or real-world data, with RCT data in treatment effect evaluation. This integration introduces unique questions regarding target population specification, causal estimand, and optimality of pooled estimators. Balancing weights have emerged as valuable tools to ensure comparability in patient characteristics, but there remains a gap in implementing them with ECs. In this study, we elucidate potential estimands of interest and propose corresponding balancing-weight-based estimators. We provide statistical and clinical definitions and interpretations of the estimands. Our extensive simulations show that different causal estimands perform differently with respect to bias and efficiency, based on the level of similarity between RCT and EC.
\end{abstract}

\section{Introduction}

Randomized controlled trials (RCTs) are often considered the gold standard for evaluating treatment effects in clinical research; however, they have several practical and ethical challenges when studying rare or life-threatening diseases. First, the limited number of eligible participants can undermine the statistical rigor of the study, leading to poor control of type I error and reduced power \citep{wu2020use}. Second, clinical equipoise (or treatment equipoise) may not be justified if inferior treatments (e.g., placebo) are assigned to participants \citep{o2020emerging,xu2020study}. Due to these challenges, many clinical trials for rare diseases tend to assign active treatments to the majority of study participants or even design a single-arm trial. Although these approaches help minimize unnecessary risks to patients, they create another challenge: evaluating the efficacy of the active treatment becomes difficult due to the lack of information on a proper control group. To address these issues, the utilization of \emph{external control} (EC) has gained prominence as a methodological approach, enabling the incorporation of control group data from independent sources \citep{viele2014use,li2023improving,swaminathan2023external,colnet2024causal}. EC data can be from control groups in historical RCTs, large registry databases, observational cohort data, or real-world data such as electronic health records (EHR).

ECs have the potential to increase statistical efficiency, improve statistical power, and reduce mean squared error (MSE) in treatment effect assessment. Additionally, external validity may be improved by including a broader patient sample. However, these benefits can only be realized with carefully selected and harmonized EC data; otherwise, the use of ECs can lead to biased and unreliable conclusions \citep{pocock1976combination, lim2018minimizing,burcu2020real}. To avoid these pitfalls, it is crucial to use EC data that closely align with the target population and the eligibility criteria of the RCT. Furthermore, outcomes in RCT and EC should be consistently defined and measured, with comparable data collection procedures (e.g., same follow-up duration). \cite{pocock1976combination} summarized acceptability conditions for the integration of RCT and EC data. These conditions require that the EC be similar to the RCT in terms of control arm treatment, eligibility criteria, treatment evaluation metric, patient characteristics, organization, and clinical investigators. Researchers often use Pocock's conditions as a guide to select valid EC sources and filter eligible EC samples \citep{swaminathan2023external}.

In recent years, there has been renewed interest in Bayesian dynamic borrowing methods for integrating ECs with trial data, including the power prior \citep{ibrahim2015power,banbeta2019modified}, commensurate prior \citep{hobbs2012commensurate} and meta-analytic predictive (MAP) prior \citep{neuenschwander2010summarizing,schmidli2014robust}. For a comparison of these methods, see \cite{viele2014use}, \cite{van2018including}, \cite{banbeta2022power} and \cite{yanchenko2023effect}. These methods have also incorporated a propensity score (PS), which is an aggregate summary of imbalance in patient characteristics between treatment arms. To achieve better control of bias, MSE, type I error, and power utilizing PS, several 2-stage PS-based Bayesian dynamic borrowing approaches have been proposed, such as the PS-integrated power prior \citep{lu2022propensity}, the PS-based MAP prior \citep{liu2021propensity}, and the PS-integrated commensurate prior \citep{wang2022propensity}.

Alternatively, classic causal inference approaches such as PS matching, PS stratification, and PS weighting can be used to integrate RCT data with EC data. While these methods are relatively easy to implement in real-world case studies, they can sometimes result in treatment effect estimates that differ from those obtained using RCT data alone~\citep{carrigan2020using, swaminathan2023external}. To address these issues, a modified PS-matching approach was proposed to identify good matches from multiple historical control groups while accounting for residual differences \citep{stuart2008matching}; further modification was proposed by \cite{yuan2019design} to reduce RCT sample size. More efficient methods, such as G-computation and doubly debiased machine learning methods, were shown to produce smaller bias and MSE compared to classic PS methods when using ECs in single-arm trials \citep{loiseau2022external}. For EC integration with two-arm traditional parallel clinical trials, \cite{li2023improving} theoretically proved the efficiency gain from utilizing ECs and proposed a doubly robust estimator. \cite{colnet2024causal} reviewed recent EC estimators that incorporate information from observational studies with RCT data to evaluate causal treatment effects.

However, the preceding methods primarily focus on the problem of estimation, while the \textit{estimand} is either never-discussed or ill-defined. When evaluating treatment effects by combining data from an EC and an RCT, it is crucial to carefully define the target population and the associated estimand, as this involves two independent populations: the RCT population and the EC population. \cite{colnet2024causal} discussed similar issues including generalizability and transportability, in the context of non-nested sampling for the RCT and observational data. In their discussion, the observational data represent a random sample from the true target population, while only the RCT employs a selective sampling mechanism. Consequently, the average treatment effect (ATE) in the target population is easily identifiable. 

In this paper, we advance the discussion by defining target estimands in scenarios where both the RCT and EC have selectively sampled from a super-population. This situation is particularly relevant in cases where the RCT is combined with EC data. Specifically, we define and compare three natural target populations for estimating treatment effects when integrating data from an RCT with an EC: (1) the population combining both RCT and EC in equal proportion to their observed sample sizes, (2) the population represented solely by the RCT, and (3) the population well-represented in both RCT and EC. We begin by defining estimands and corresponding estimators for these three target populations, exploring their practical interpretations and implications (Sections~\ref{sec:problem_setting}-\ref{sec:estimator}). We focus on propensity score weighting estimators in this this paper, although other estimators can be constructed for the estimands of interest.  Following this, we conduct extensive simulations (Section~\ref{sec:simulation}) to evaluate the bias and MSE of the estimators across various data-generating scenarios, including different sample size ratios between CC and EC, varying levels of heterogeneous treatment effects (HTE), and different degrees of covariate imbalance between the RCT and EC. Finally, we provide further discussions on the application of ECs in clinical research (Section \ref{sec:discussion}).
The main question to be answered in the paper is: For a given estimand, how effectively can it be estimated using propensity score weights?

\section{Problem Setting}\label{sec:problem_setting}

\subsection{Sampling Scheme}
In most clinical research integrating RCT with EC data, a non-nested sampling design where RCT and EC data are collected separately, each with unknown sampling probabilities \citep{dahabreh2021study, colnet2024causal}, is often considered.
Each dataset is harmonized with respect to inclusion criteria and represents a clinically relevant target population, but none is randomly sampled from ``the" true target population. In fact, ``the" true target population is not explicitly clear. In contrast, a nested design would involve randomly sampling from a well-defined population and then allocating participants to the trial versus external data, resulting in a mutually exclusive partition of RCT and EC patients \citep{dahabreh2021study}. Since such nested designs are rare, our focus here is on the non-nested design. 

Defining clear causal estimands starts with a conceptual model for the target population and samples drawn from it. Suppose there exists an unknown super-population, $\Omega$, which contains all patients meeting broad clinical inclusion/exclusion criteria within an eligible setting (e.g. country, inpatient vs. outpatient, etc.), regardless of whether they can feasibly be sampled. Within this super-population, patients may be prioritized differently based on their characteristics and the research priorities. For example, international clinical trials typically over-represent certain regions or countries that are prioritized by regulators, rather than sample according to population sizes. This is purposeful. Even if the super-population of eligible patients could be enumerated in a sampling frame, it would not be desirable to view this as ``the" target population.  Instead, the super-population contains many candidate sub-populations that meet criteria for scientific relevance, each refined by operational priorities.

In designing a study, we select ``a" target population based on practical decisions and then sample from it.  Let $\Omega_1, \ldots,\Omega_\infty$ be all sub-populations of $\Omega$ that meet minimal criteria for clinical relevance, i.e. the potential target populations. These minimal criteria are typically defined by inclusion/exclusion criteria, shaped by investigators, regulatory bodies, journal expectations, or community expectations. Each potential target population, $\Omega_p$ with $p\in\{1,2,\ldots\}$, varies due to differences in unconstrained features such as country, site selection, site enrollment rate, and patient enrollment preferences. These potential target populations, $\Omega_p$'s, are not mutually exclusive and may overlap substantially. The set of sampling processes that give rise to $\Omega_1, \ldots,\Omega_\infty$ are denoted as $p_1(x),\ldots,p_\infty(x)$, respectively. We can think of $p_1(x),\ldots,p_\infty(x)$ as a set of clinically relevant sampling processes. The samples we observe as RCT and EC are products of their respective unknown sampling processes, e.g. $p_1(x)$ for the RCT and $p_2(x)$ for the EC, combined with random sampling variability.

\begin{figure}[ht!]
    \begin{center}
    \subfigure[Sampling scheme]{
        \includegraphics[scale=0.5]{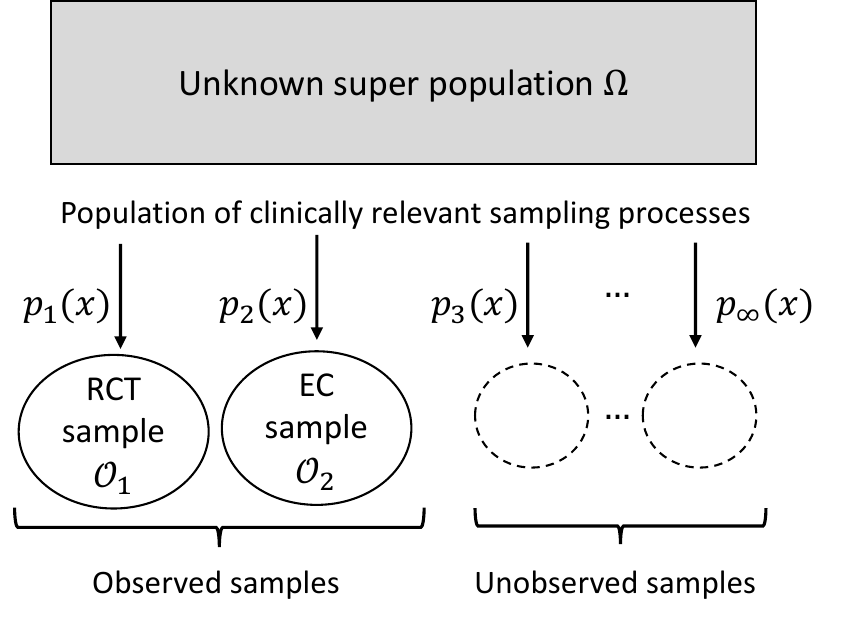}
    } 
    \subfigure[Target population distribution]{
        \includegraphics[scale=0.5]{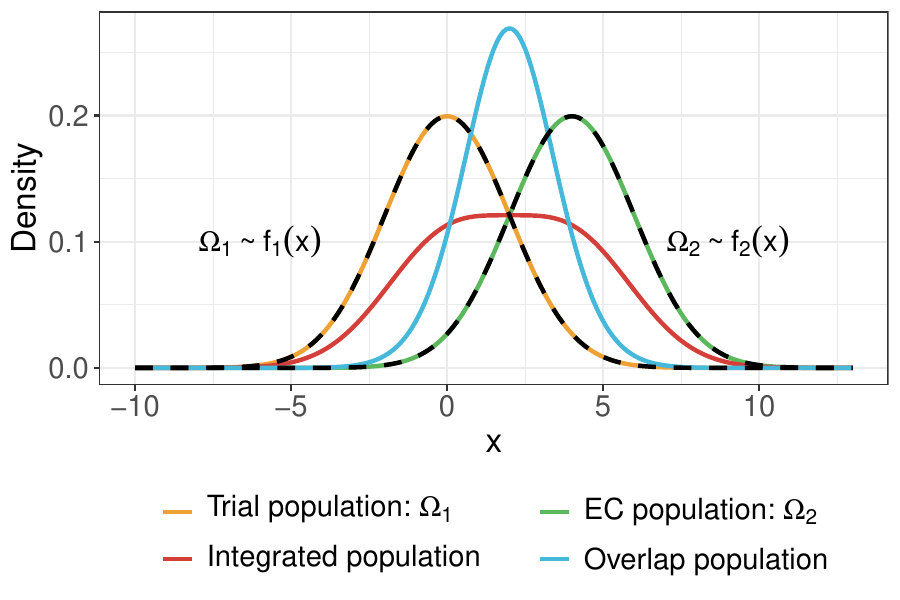}
    }
    \end{center}
    \caption{Illustrations of the problem setting. (a) Schematic of the non-nested, non-random sample selection processes that generate the RCT and EC samples. (b) Covariate density curves of target populations of interest, modified Figure 1 in \cite{li2018balancing}. $f_1(x)$ and $f_2(x)$ are densities for $N(0,2^2)$ and $N(4,2^2)$, respectively. Integrated population density $f_I=0.5f_1(x)+0.5f_2(x)$, $f_{\text{trial}}\propto\pi(x) f_I$, $f_{\text{EC}}\propto(1-\pi(x)) f_I$ and $f_{\text{overlap}}\propto\pi(x)(1-\pi(x))f_I$ with $\pi(x)=\Pr(Z=1|X=x)$.}
    \label{fig:process}
\end{figure}

\subsection{Observed Data Structure}
The observed data structure is summarized in Table~\ref{tb:data_structure}. The RCT sample, denoted by $\cO_1$, consists of $N_1$ participants, with $N_{11}$ randomized to the treatment arm and $N_{10}$ to the control arm (i.e., $N_{11} + N_{10} = N_1$). Without loss of generality, the unknown RCT sampling process is denoted as $p_1(x)$, representing population $\Omega_1$. The EC data, denoted by $\cO_2$, contains $N_2$ participants sampled according to unknown process, $p_2(x)$, representing population $\Omega_2$. The indices assigned to these two populations are arbitrary and could represent any two processes. As a result, the total number of observed subjects is $N=N_1+N_2$. 

Potential outcomes are defined as $Y(a)$, where $a\in\{0,1\}$, with $Y(1)$ representing the potential outcome under treatment and $Y(0)$ representing the potential outcome under control. Let $Y$ denote the observed outcome of interest, $A$ denote the treatment assignment ($A=1$ for patients receiving treatment, $A=0$ for patients receiving control), and $Z$ denote the data source ($Z=1$ for patients from the RCT, $Z=0$ for patients from the EC). Let $X$ denote covariates with density functions $f_{1}(x)$ and $f_{2}(x)$ corresponding to samples from populations $\Omega_1$ and $\Omega_2$, respectively. For every patient, we observe the set $\{Y,X,A,Z\}$.

\begin{table}[ht!]
\centering
\caption{Data structure of RCT and EC. $\checkmark$ and ? represent observed or unobserved, respectively, modified from Table 1 in \cite{li2023improving}.}\label{tb:data_structure}
\begin{threeparttable}
\begin{tabular}{cccccccc}
\toprule
       &                     & Data Source & Treatment & \multicolumn{2}{c}{Potential Outcome} & Observed Outcome & Covariates\\ 
       &                     & $Z$      & $A$       & $Y(1)$            & $Y(0)$            & $Y$              & $X$\\\midrule
$\cO_1$ (RCT) & 1                   & 1        & 1         & $\checkmark$      & ?                 & $\checkmark$     & $\checkmark$\\
       & $\vdots$            & $\vdots$ & $\vdots$  & $\vdots$          & $\vdots$          & $\vdots$         & $\vdots$\\
       & $N_{11}$            & 1        & 1         & $\checkmark$      & ?                 & $\checkmark$     & $\checkmark$\\\cdashline{2-8}
       & $N_{11}+1$          & 1        & 0         & ?                 & $\checkmark$      & $\checkmark$     & $\checkmark$\\
       & $\vdots$            & $\vdots$ & $\vdots$  & $\vdots$          & $\vdots$          & $\vdots$         & $\vdots$\\
       & $N_1$     & 1        & 0         & ?                 & $\checkmark$      & $\checkmark$     & $\checkmark$\\\hline
$\cO_2$ (EC) & $N_1+1$             & 0        & 0         & ?                 & $\checkmark$      & $\checkmark$     & $\checkmark$\\
       & $\vdots$            & $\vdots$ & $\vdots$  & $\vdots$          & $\vdots$          & $\vdots$         & $\vdots$\\
       & $N$& 0        & 0         & ?                 & $\checkmark$      & $\checkmark$     & $\checkmark$\\ \bottomrule
\end{tabular}
\end{threeparttable}
\end{table}

\section{Causal Estimands}\label{sec:estimand}

The average treatment effect (ATE), defined as $\E[Y(1)-Y(0)]$, is a standard causal estimand. However, this notation does not explicitly state the population over which the expectation is taken. It is implied that the population is unequivocal.  In the context of RCT augmented by EC data, there are multiple ways to combine the samples $\cO_1$ and $\cO_2$, leading to various potential target populations and corresponding estimands.

Naively combining the RCT and EC samples in proportion to their sizes yields a mixture. Using a slight abuse of notation, we denote the infinite population corresponding to the mixture of $\cO_1$ and $\cO_2$ in proportion to their sample sizes as $\Omega_I=\Omega_1\cup\Omega_2$. The mixture covariate density function is 
\begin{equation}
    f(x) = \lambda f_{1}(x) + (1-\lambda) f_{2}(x),
\end{equation}
where $\lambda= N_1/N$ \citep{li2023improving}. $\Omega_I$ is referred to as the ``integrated population'' with density function $f(x)$, which is a weighted average of covariate densities from $\Omega_1$ and $\Omega_2$. The corresponding average treatment effect estimand is $\E_f[Y(1)-Y(0)]$, where the expectation is with respect to $f(x)$.

More generally, we can define the ATE over $g$ as $\E_g[Y(1)-Y(0)]$, where $g$ represents the target population density. Special choices of $g$ define the estimands in Section \ref{Est}.

\subsection{Choice of Target Population}

Four target populations are of particular interest including the RCT population, the EC population, a mixture according to sample size, and a weighted mixture. Recent publications have emphasized either the RCT population \citep{li2023improving} or the EC population \citep{colnet2024causal} as the target population. The RCT population is prioritized when the goal is to borrow efficiency from external data without altering the interpretation of the current randomized trial. Inference is restricted to patients like those enrolled in the RCT (often quite narrow). In contrast, the EC population is naturally prioritized when the EC data represents a broad, clinically generalizable population \citep{colnet2024causal}. 

Figure~\ref{fig:process} (b) exemplifies these populations with respect to a single covariate $X$. The RCT population (represented by the yellow density line) tends to have lower values of covariate $X$. In contrast, the EC population (represented by the green density line) tends to have high values of covariate $X$. These two populations could have substantially different average causal effects.  For example, suppose that the treatment effect increases as $X$ decreases. In this case, the RCT population will have more people with large treatment effects and a larger average causal effect. 

The ``integrated population''(represented by the red density line) is implicitly assumed when RCT and EC data are naively combined. The mixture density $f(x)$ looks like an even combination of $f_1(x)$ and $f_2(x)$. However, viewing this as ``the" target population can be questionable, as the mixture proportion depends on the observed sample sizes. If the EC is large it may dominate the integrated population.  When sample sizes are a result of convenience and not quality, this is not desirable. 

Additionally, we are interested in an ``overlap population'' (represented by the blue density line) that is a weighted combination of RCT and EC. This overlap population is motivated by the propensity score literature for treatment comparisons. In that context, it emphasizes a population for whom the treatment decision is closest to equipoise (uncertain).  In this context, it includes all units in $\Omega_1\cup\Omega_2$, but assigns greater weight to individuals whose joint covariate values are more likely in both $\Omega_1$ and $\Omega_2$ \citep{li2018balancing}, that is people who appear consistently in both samples. In Figure~\ref{fig:process} (b) we see that the overlap population resembles the integrated population but with smaller tails and greater density in the region of covariate overlap. Some covariate values from the RCT are not represented in the overlap population at all, e.g. $X=-4$. 

In the non-nested setting, any of these populations could be clinically relevant. Depending on the nature of the RCT and EC, some might be preferred. Baseline characteristics corresponding to each target population can be displayed to assist in determining the suitable target population. Beyond theoretical properties, the actual data can be used to inform clinical relevance.

\subsection{Estimands}
\label{Est}
To define the causal treatment effect estimands, we make use of a PS. In the context of non-nested design considering RCTs integrated with EC data, we define a PS that summarizes the differences \textit{between RCT and EC populations}, denoted by $\pi(x)$, as follows: 
\begin{equation}\label{eq:def pi(x)}
\begin{aligned}
    \pi(x) &\equiv\Pr(Z=1|X=x).
\end{aligned}
\end{equation}
It is important to note that this PS differs from the treatment PS, $e(x)=\Pr(A=1|X=x)$, commonly used in nested designs to summarize differences between patients assigned to treated and untreated groups \citep{rosenbaum1983central}. In the context of EC integration, we require a summary score that captures differences between the two data sources (RCT and EC), rather than differences arising from the treatment assignment mechanism. The PS $\pi(x)$ has the following relationship with $\lambda$, $f_1(x)$ and $f_2(x)$:
\begin{equation}\label{eq:pi(x)}
\begin{aligned}
    \pi(x) &= \frac{\Pr(X|Z=1)P(Z=1)  }{\Pr(X|Z=1)P(Z=1) + \Pr(X|Z=0)P(Z=0) } \\
    &= \frac{f_1(x)\lambda  }{f_1(x)\lambda + f_2(x)(1-\lambda) } \\
    &= \frac{f_1(x)\lambda  }{f(x)}.
\end{aligned}
\end{equation}

\begin{table}[h]
\centering
\caption{Target population and corresponding balancing weights.}\label{tb:weights}
\begin{tabular}{ccccc}
\toprule
\multirow{2}{*}{Target Population} & \multirow{2}{*}{$h(x)$} & \multirow{2}{*}{Estimand}  & Weight \\
&&& $(w_1(x),w_0(x))$  \\\midrule
\vspace{3mm}RCT & $\pi(x)$ & ATT & $\left(1,\frac{\pi(x)}{1-\pi(x)}\right)$         \\
\vspace{3mm}EC & $(1-\pi(x))$ & ATEC & $\left(\frac{1-\pi(x)}{\pi(x)},1\right)$ \\
\vspace{3mm}Integrated  & 1 & ATI & $\left(\frac{1}{\pi(x)},\frac{1}{1-\pi(x)}\right)$ \\
Overlap  & $\pi(x)(1-\pi(x))$ & ATO & $\left(1-\pi(x),\pi(x)\right)$ \\\bottomrule  
\end{tabular}
\end{table}

Following \cite{li2018balancing}, the ATE estimand over covariate distribution $g(x)$ can be written as
\begin{equation}\label{eq:tau_g}
    \tau_g \equiv \E_g[Y(1)-Y(0)] \equiv \frac{\int \tau(x)h(x)F(\dd x)}{\int h(x)F(\dd x)},
\end{equation}
where $\tau(x) = \mu_1(x)-\mu_0(x) = \E[Y(1)-Y(0)|X=x]$ is the conditional average treatment effect (CATE), $F(x)$ is the cumulative distribution function (CDF) of the integrated population, and $h(x)$ is a tilting function defined such that $g(x) = h(x)f(x)$. The tilting functions, $h(x)$, that correspond to each population are shown in Table~\ref{tb:weights}. The average treatment effect among the RCT (trial) population (ATT) is
\begin{equation}
    \tau_\ATT\equiv \frac{\int\tau(x)\pi(x)F(\dd x)}{\int\pi(x)F(\dd x)} = \E_{f_1}[Y(1)-Y(0)],
\end{equation}
and the average treatment effect among the EC population (ATEC) is
\begin{equation}
    \tau_\text{ATEC}\equiv \frac{\int\tau(x)(1-\pi(x))F(\dd x)}{\int(1-\pi(x))F(\dd x)} = \E_{f_2}[Y(1)-Y(0)],
\end{equation}
and the average treatment effect among the integrated population (ATI) is
\begin{equation}\label{eq:pate1}
    \tau_\ATI\equiv \int \tau(x) F(\dd x) = \E_f[Y(1)-Y(0)].
\end{equation}
The average treatment effect among the overlap population (ATO) is
\begin{equation}\label{eq:pato1}
    \tau_\ATO\equiv\frac{\int\tau(x)\pi(x)(1-\pi(x))F(\dd x)}{\int\pi(x)(1-\pi(x))F(\dd x)}.
\end{equation}

Up to this point, we include the ATEC estimand for completeness. A large body of literature on trial generalizability has addressed the ATEC \citep{lee2023improving,colnet2024causal}. We therefore do not focus on this estimand subsequently.

\section{Identification and Estimation}\label{sec:estimator}

\subsection{Assumptions for Identifiability of Estimands}
We borrow \cite{rosenbaum1983central}'s causal framework with potential outcomes. Assumption~\ref{ass:sutva} summarizes the underlying assumptions of the data structure in Table~\ref{tb:data_structure}, that is, the standard stable unit treatment value assumption (SUTVA).
\begin{assumption}[SUTVA]\label{ass:sutva}
$Y=AY(1)+(1-A)Y(0)$.
\end{assumption}
To identify the causal treatment effect estimands for the integrated, RCT, and overlap populations, additional assumptions are required \citep{li2023improving}:
\begin{assumption}[Strong Ignorability for Treatment within RCT]\label{ass:ignore_trt}
~
\begin{enumerate}[label=(\roman*)]
    \item Unconfoundedness: $\{Y(1),Y(0)\}\perp A \mid X, Z=1$;
    \item Overlap: $0<\Pr(A=1|X=x,Z=1)<1$ for all $x$ with $\Pr(X=x|Z=1)>0$.
\end{enumerate}
\end{assumption}
\begin{assumption}[Mean Exchangeability for $Y(1)$ and $Y(0)$ between RCT and EC]\label{ass:mean_exchange}
$\E[Y(1)|X,Z=1]=\E[Y(1)|X,Z=0]$ and
$\E[Y(0)|X,Z=1]=\E[Y(0)|X,Z=0]$.
\end{assumption}
\begin{assumption}[Overlap between RCT and EC]\label{ass:overlap}
$0<\Pr(Z=1|X)<1$.
\end{assumption}

Under the assumption of purely random treatment assignment in RCT, $A$ is independent of $X$, i.e.,
\begin{equation*}
    \Pr(A=a|X,Z=1)=\Pr(A=a|Z=1),
\end{equation*}
where $a\in\{0,1\}$. The assumption of mean exchangeability is commonly used in literature on RCT augmented by EC \citep{li2023improving,colnet2024causal}. It is closely connected to the idea of unmeasured confounding, which may be more familiar to some.  A slightly stronger set of assumptions is that (1) there are no unmeasured confounders that differ between RCT and EC subjects and that (2) being in the RCT itself has no causal effect on controls (no Hawthorne effects) \citep{li2023improving}.  Mean exchangeability includes both assumptions with respect to the mean potential outcomes. 

To identify ATT, the mean exchangeability assumption only needs to hold for $Y(1)$, allowing Assumption~\ref{ass:mean_exchange} to be relaxed. For ATI and ATO, however, the mean exchangeability must hold for both $Y(1)$ and $Y(0)$. Detailed identification derivations are provided in Appendix~\ref{app:identify}.

\subsection{Estimators}
Enlightened by weighting identifications of ATI, ATT and ATO, we propose the following IPW estimators for ATI, ATT, and ATO:
\begin{equation}\label{eq:estimator}
\hat\tau = \frac{\sum_{i=1}^N w_1(X_i)A_iZ_iY_i}{\sum_{i=1}^N  w_1(X_i)A_iZ_i} - \frac{N_{10}}{N_{10}+N_2}\frac{\sum_{i=1}^N w _1(X_i)(1-A_i)Z_iY_i}{\sum_{i=1}^N w_1(X_i)(1-A_i)Z_i} - \frac{N_{2}}{N_{10}+N_2}\frac{\sum_{i=1}^N w_0(X_i)(1-A_i)(1-Z_i)Y_i}{\sum_{i=1}^N w_0(X_i)(1-A_i)(1-Z_i)},
\end{equation}
where $w_1(X_i)$ and $w_0(X_i)$ are the weights defined in Table~\ref{tb:weights}. For ATI, $w_1(x)=1/\pi(x)$ and $w_0(x)=1/(1-\pi(x))$; for ATT, $w_1(x)=1$ and $w_0(x)=\pi(x)/(1-\pi(x))$; and for ATO, $w_1(x)=1-\pi(x)$ and $w_0(x)=\pi(x)$. In practice, when true $\pi(x)$ are unknown, one can fit a PS model (e.g., logistic regression) to estimate $\pi(x)$, and then plug the estimated PS, $\widehat{\pi}(x)$, into $w_1(x)$ and $w_0(x)$.

Table~\ref{tb:weights} shows the raw weights, which are typically normalized so that they sum to one within each treatment group. For example, the estimator $\hat\tau$ includes this normalization step in each term's denominator.  

\subsection{Computation}
Simulations and data analysis were performed in R \citep{R}. To estimate the PS, we utilized the \texttt{PSweight} package \citep{zhou2020psweight}. In the simulations, a simple logistic regression model was employed to estimate the PS (details are provided in Section~\ref{sec:simulation}). The R code used for the simulation is available at \url{https://github.com/Peijin-Wang/EC_Estimator}.

\section{Simulation}\label{sec:simulation}
Our simulation study evaluates the performance of estimators for ATI, ATT, and ATO under various settings: (1) different degrees of similarity between RCT and EC, (2) different levels of HTE, and (3) different sample sizes between RCT and EC. 

\subsection{Simulation Setup}
We generate the potential outcomes using the following linear models: 
\begin{equation}\label{eq:outcome_model}
\begin{aligned}
    Y_i(0) &= \beta_0+\beta_1X_{1i} + \beta_2X_{2i} + \varepsilon_i,\\
    Y_i(1) &= \beta_0+\beta_1X_{1i} + \beta_2X_{2i} + \beta_{\text{trt}} + \phi_1 X_{1i} + \phi_2X_{2i} + \varepsilon_i,
\end{aligned}
\end{equation}
where $X_{1i}$ is a binary covariate, $X_{2i}$ is a continuous covariate, and $\varepsilon_i {\sim} N(0,1)$ is random error. RCT and EC data are independently generated using Eq~\eqref{eq:outcome_model}. 

For RCT, $X_{1i}\sim\Bernoulli(0.5)$ and $X_{2i}\sim N(0,1)$. For EC, we vary the distribution of $X_{2i}$ to simulate different scenarios of similarity between RCT and EC, while $X_{1i}\sim\Bernoulli(0.5)$ remains the same as RCT. We consider eight different distributions of $X_{2i}$ in EC:
\begin{itemize}
    \item EC1: $N(0,1)$, no difference in the distribution of $X_2$ between EC and CC;
    \item EC2 to EC4: $N(0.5,1)$, $N(1,1)$, $N(2,1)$, mean shift in the distribution of $X_2$ between EC and CC;
    \item EC5: $N(0,1.5)$, distribution of $X_2$ is more variable in EC;
    \item EC6 to EC8: $N(0.5,1.5)$, $N(1,1.5)$, $N(2,1.5)$, mean shift with high variability in the distribution of $X_2$ in EC.
\end{itemize}
The first rows of Figures~\ref{fig:x2_1} and~\ref{fig:x2_2} in Appendix \ref{app:add_sim_results} visualize distributions of $X_2$ between RCT and EC.

Table~\ref{tb:coef_setting} summarizes the true parameter settings for HTE and sample sizes. In Eq~\eqref{eq:outcome_model}, we set the treatment effect, $\beta_{\text{trt}}$, to zero, and consider three types of HTE: no HTE ($\phi_1=\phi_2=0$), moderate HTE ($\phi_1=\phi_2=0.25$), and large HTE ($\phi_1=\phi_2=0.5$). Additionally, under each HTE setting, we vary the RCT treatment-control allocation ratio and RCT-EC sample size ratio. First, we consider 1:1 (settings 1-9) and 3:1 (settings 10-18) treatment-control allocation ratio within RCT ($N_1=200$). The 3:1 allocation ratio reflects a practical situation where fewer RCT participants are assigned to the control group due to the leverage of information from EC. Second, we consider three different CC-EC ratios: 1:1, 1:3, and 1:10. The CC-EC ratio of 1:10 reflects a situation where EC is from a large real-world dataset.

Given the simulated potential outcomes, the observed outcome $Y_i$ is calculated as follows
\begin{equation}
    Y_i = A_iY_i(1) + (1-A_i)Y_i(0).
\end{equation}
To sum up, we consider 144 simulation scenarios (8 distributions of $X_2$ in EC $\times$ 18 outcome models). For each scenario, we simulate 1000 pairs of RCT and EC.

\subsection{True Estimand Calculation and Simulation Evaluation Metrics}
Within each simulation setting, we compute the true values of the estimands based on the true CATE, defined as $\tau(x) =\mu_{0}(x) - \mu_{1}(x) = \beta_{\text{trt}} + \phi_1 x_{1} + \phi_2x_{2},$ and the true PS $\pi(x)$, defined as in Eq~\eqref{eq:pi(x)}, where $f_1(x)$ is the $(X_1,X_2)$ joint density of RCT, $f_2(x)$ is the $(X_1,X_2)$ joint density of EC, and the mixture proportion $\lambda=N_1/(N_1+N_2)$ (see Table~\ref{tb:coef_setting}). The true population estimands ATI/ATT/ATO can be computed using the formulas in Eq~\eqref{eq:pate1} to \eqref{eq:pato1}. Monte Carlo integration is employed to numerically compute the integrals. Detailed procedures for calculating the true value of the estimands are provided in Appendix~\ref{app:compute_estimand}.

It is important to note, as shown in Table~\ref{tb:app_sim_estimand}, that the true estimand values differ across settings, except when no HTE exists ($\phi_1=\phi_2=0$). The true values of the ATI and ATO estimands vary across all settings, while the true values of the ATT estimand change only with variations in the HTE setting. This is because the target populations for ATI and ATO estimands contain all units in the integrated mixture of the RCT and EC populations, with the joint covariate distribution depending not only on the RCT density $f_1(x)$ and the EC density $f_2(x)$ but also on the mixture proportion $\lambda$. Consequently, the target populations for ATI and ATO change under different parameter settings and ECs. In contrast, the target population for the ATT estimand is the RCT population, independent of the EC population, so the true value of the ATT estimand is influenced only by $\phi_1$ and $\phi_2$.

For each simulated pair of RCT and EC, we estimate $\pi(x)$ by fitting logistic regression $\logit(\pi(x))\sim X_1+X_2$, and then compute the proposed estimators as defined in Eq~\eqref{eq:estimator}. The performance of those estimators is evaluated using bias and MSE defined as follows:
\begin{equation}
\begin{aligned}
    &\text{Bias}\equiv B^{-1}\sum_{b=1}^B(\hat\tau_{kb} - \tau_k),\\
    &\text{MSE}\equiv B^{-1}\sum_{b=1}^B(\hat\tau_{kb} - \tau_k)^2,
\end{aligned}
\end{equation}
where $\hat\tau_{kb}$ is the estimator obtained at the $b^{\text{th}}$ iteration, $\tau_k$ is the true estimand for $k\in\{\ATI,\ATT,\ATO\}$ and $B=1000$. Importantly, each estimator is compared to its own estimand. Thus, bias and variance are not attributable to differences in the intended target population. In many cases, each of the well-known target populations is clinically relevant and could be of interest.

\subsection{Simulation Results}
We present the simulation results for the 1:1 treatment-control allocation ratio (settings 1-9) because they are similar to those obtained under the 3:1 treatment-control allocation ratio (settings 10-18). The results for settings 10-18 are provided in Appendix~\ref{app:add_sim_results}.

Figure~\ref{fig:bias1} shows the bias under settings 1-9 (detailed bias values are provided in Table~\ref{tb:app_sim_bias}). Overall, the ATO estimator demonstrates almost unbiased results across all EC and HTE settings. The ATT estimator generally exhibits a smaller bias compared to the ATI estimator in most settings. The ATI estimator tends to struggle significantly when the RCT and EC populations differ, when HTE is large, or when sample sizes of the RCT and EC are imbalanced. In cases where the RCT and EC populations have relatively similar distributions of $X_2$ (EC1, EC2, EC3, and EC5), all three estimators show almost no bias. However, as the EC population becomes more distinct from the RCT population, the performance of the estimators differentiates. For example, when integrating EC4 ($X_2\sim N(2,1)$) with RCT data ($X_2\sim N(0,1)$) under setting 1 (no HTE, $N_{10}:N_2=1:1$ and $N_2=100$), the bias is -0.13, -0.10, and 0.01 for the ATI, ATT and ATO estimators, respectively (see Table~\ref{tb:app_sim_bias}). This trend becomes more pronounced as HTE increases (from 0 to 0.5) and as the EC sample size increases (moving from left to right sub-panels within each EC case). Across all ECs, as the mixture proportion $\lambda$ decreases (or equivalently, as $N_2$ increases), the bias of the ATI estimator increases, while the ATO and ATT estimators tend to provide more consistent results. An exception occurs in EC8, where the ATT estimator's bias slightly increases as $N_2$ increases. 

\begin{figure}[htbp]
    \centering
    \includegraphics[width=\linewidth]{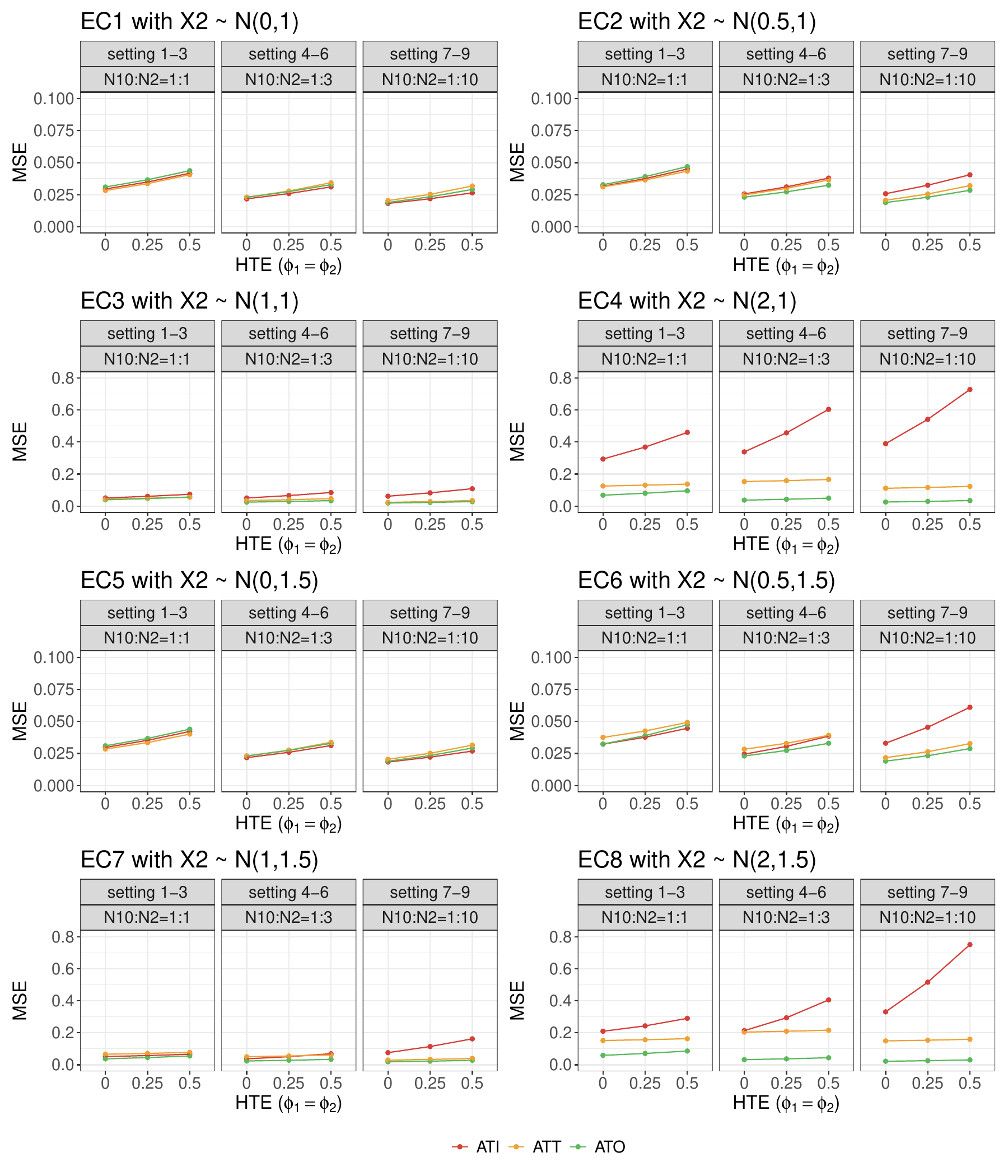}
    \caption{MSE of proposed estimators in estimating ATI, ATT and ATO under 1:1 treatment-control allocation ratio of RCT (settings 1 to 9).}
    \label{fig:mse1}
\end{figure}

Figure~\ref{fig:mse1} illustrates the MSE under settings 1-9. The MSE results follow similar patterns to those observed for bias. Overall, the ATO estimator exhibits the smallest MSE, followed by the ATT and ATI estimators. When the RCT and EC populations are similar (EC1, EC2, EC3, and EC5), all estimators demonstrate low MSE. However, the ATI estimator shows significantly higher MSE with EC4 and EC8 especially when HTE is large. 

Figures~\ref{fig:x2_1} and~\ref{fig:x2_2} exhibit the weighted distribution of $X_2$. For EC1, EC2, EC5, and EC6, the weighted distributions of $X_2$ are similar across the three estimands. However, when the EC and RCT populations have more distinct $X_2$ distributions (EC3, EC4, EC7, and EC8), the weighted distributions differ across estimands. The ATI weights lead to the most unstable weighted distributions, while the ATT weights produce weighted distributions that closely resemble the RCT distribution. The ATO weights yield weighted distributions of $X_2$ that fall between the distributions of the RCT and EC, reflecting the overlapping population.

\section{Discussion}\label{sec:discussion}
In this paper, we consider various target populations and potential estimands under a non-nested design involving RCT and EC, utilizing the tilting function and balancing weight framework proposed by \cite{li2018balancing}. Different average causal estimands, including ATI, ATT, and ATO, are defined and interpreted. Enlightened by the identification formula, IPW estimators are proposed for each estimand. Simulation results indicate that when there is a mild difference between RCT and EC, all three estimators perform similarly. With a moderate difference, both ATT and ATO estimators demonstrate well-controlled bias and MSE. However, with extreme differences, the ATO estimator is the best. Data analysis reveals that when extreme PSs are presented, the ATI and ATT estimators may lose efficiency.

The primary context for which the ATI and ATT estimators perform poorly is when the EC deviates substantially from RCT. For the ATI estimator, units in both RCT and EC may have extreme weights, while for the ATT estimator, units in EC may have extreme weights. Consequently, responses corresponding to extreme weights will dominate the estimator, potentially leading to more biased estimates. The benefit of the ATO estimator is that it avoids assigning extremely large weights to units, resulting in more stable bias and MSE regardless of how EC deviates from RCT. Extreme PSs indicate the lack of overlapping information in the covariate space between the studies. Any estimand that includes this sparse covariate space within the target population will be harder to estimate. These findings are analogous to the standard PS setting as discussed by \cite{li2019addressing}.

Although the preceding findings are expected, their implications in the current context are noteworthy. In nested designs, it is common to assume that the sample is representative of ``the'' target population, which is then partitioned into treatment and control groups.  As such, caution is needed when interpreting causal estimand that deviates from this sampled population \citep{li2019addressing}. In contrast, the non-nested design used here provides valuable insights from each sample regarding the target population. Multiple samples, which share common inclusion/exclusion criteria but differ in practical aspects, contribute to a more complete understanding of the underlying population. In this case, a combination of the RCT and EC samples may be better than either one.  While the integrated population is attractive, it depends on arbitrary sample sizes. The sampling mechanism that yields the largest sample sizes is not necessarily the most important. Any weight $\lambda$ could be used to create an intentional mixture. Alternatively, focusing on the overlap population highlights the shared information across multiple samples.  The overlap population and the ATO estimand would naturally arise by asking: First, \textit{Which population is well represented by multiple clinically relevant samples?} Second, \textit{What is the treatment effect in that population?}

Regardless of the estimand, it is essential to describe the weighted samples. The baseline characteristics table serves as a useful tool for making inferences about the target population, with or without weights.  Researchers need to give careful attention to the baseline characteristics of the population under study and note that the treatment effect conclusions apply only to that specific population. 

The proposed estimators have several limitations. First, the performance of the ATI and ATT estimators is highly dependent on the degree of overlap between RCT and EC. To assess this overlap, one efficient way is to plot the PS histogram and compare the weighted outcomes of CC and EC. When there are no extreme PS and weighted outcomes of CC and EC are similar, the proposed estimators can be applied with greater confidence. As RCT and EC have different sampling processes, it is difficult to ensure good performance of the ATI estimator, making the ATT estimator a more reliable option. Second, when unmeasured confounders are present and the mean-exchangeability assumption does not hold, all of the proposed estimators are at risk for bias. A potential solution to this issue is to account for the outcome similarity between CC and EC in the estimation procedure. For example, \cite{gao2023integrating} proposed a data-adaptive integrative framework to estimate treatment effect while considering the outcome similarity.

While this work focuses on propensity score weighted estimators, the findings have implications to other domains. For example, augmented estimators and some Bayesian dynamic borrowing methods also incorporate a propensity score. These estimators too, can be designed to target the different estimands outlined here. Similar performance characteristics might observed, as the overlap population estimand is in some ways easier to estimate efficiently. Different approaches to estimation may ultilize that same phenomenon, When it is clinically relevant to permit different estimands.

\section{Acknowledgement}
This work was funded by Burroughs Wellcome Fund.

\newpage

\begin{figure}[htbp]
    \centering
    \includegraphics[width=\linewidth]{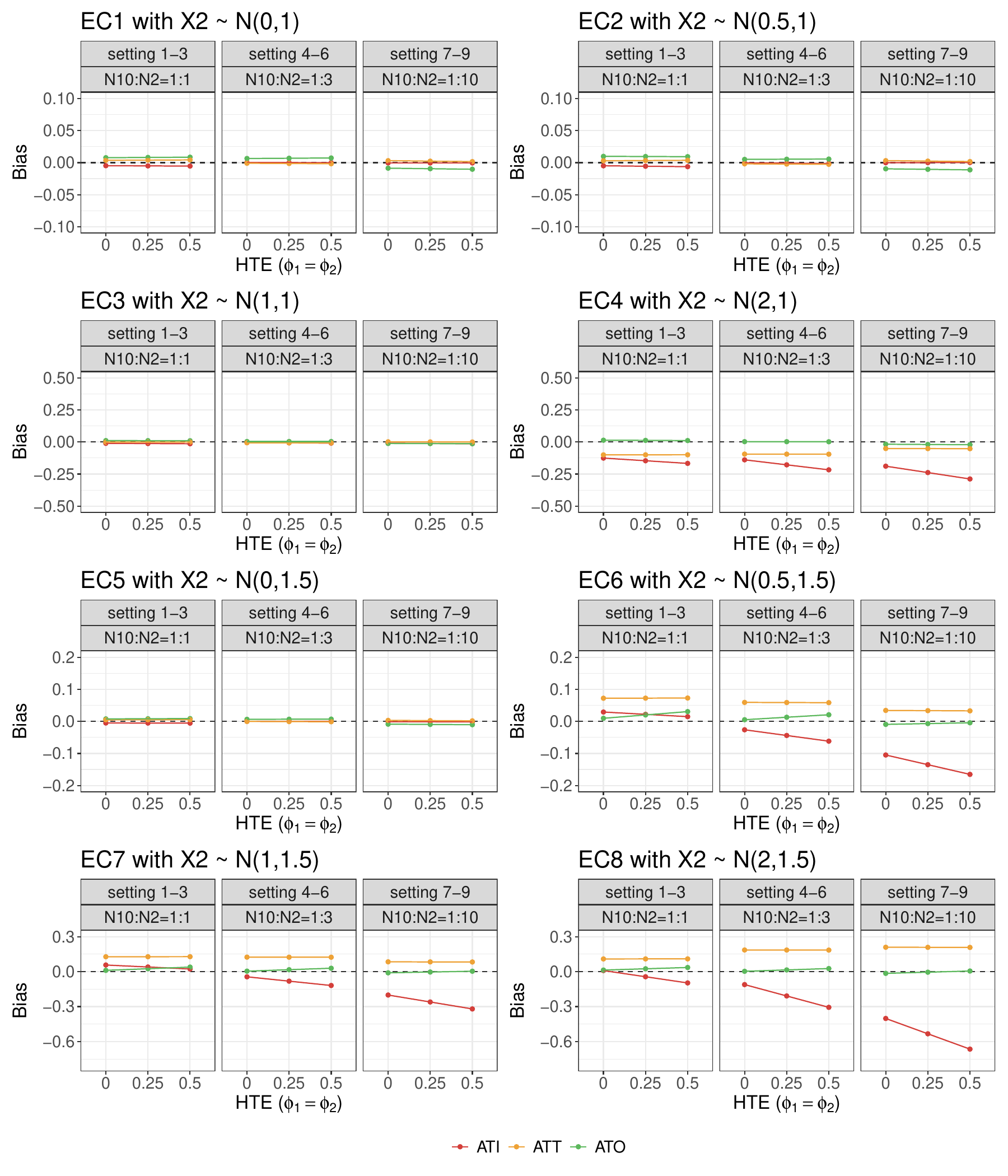}
    \caption{Bias of proposed estimators in estimating ATI, ATT and ATO under 1:1 treatment-control allocation ratio of RCT (settings 1 to 9).}
    \label{fig:bias1}
\end{figure}

\newpage

\numberwithin{equation}{section}
\numberwithin{table}{section}
\numberwithin{figure}{section}

\begin{appendices}
\section{Estimand Identification}\label{app:identify}
Using balancing weights, estimands of interest can be defined using CATE and tilting function as shown in Eq~\eqref{eq:tau_g}. We can rewrite $\tau_g$ as follows by replacing integrals with expectations:
\begin{equation}
    \tau_g = \frac{\E[\tau(X)h(X)]}{\E[h(X)]}.
\end{equation}
Let's start with identifying CATE $\tau(x) = \E[Y(1)-Y(0)|X=x]$. $\tau(x)$ is defined as
\begin{equation}\label{eq:app_cate}
\begin{aligned}
\tau(x)&=\E[Y(1)-Y(0)|X=x]\\
&= \E[Y(1)-Y(0)|X=x,Z=1]\Pr(Z=1|X=x) + \E[Y(1)-Y(0)|X=x,Z=0]\Pr(Z=0|X=x).
\end{aligned}
\end{equation}
In this paper, treatment arm information comes from RCT only, and control arm information comes from both RCT and EC. In other words, $Y(1)$ should be identified using treatment arm units in RCT and $Y(0)$ can be identified using units from CC and EC. Under Assumption~\ref{ass:ignore_trt}, $\E[Y(1)-Y(0)|X,Z=1]$ can be written as
\begin{equation}
\begin{aligned}
\E[Y(1)-Y(0)|X,Z=1]&=\E[Y(1)|X,Z=1]-\E[Y(0)|X,Z=1]\\
&= \E[Y(1)|X,A=1,Z=1]-\E[Y(0)|X,Z=1] \\
&= \E[Y|X,A=1,Z=1] - \E[Y(0)|X,Z=1],
\end{aligned}
\end{equation}
where $\E[Y(0)|X,Z=1]$ can be identified as
\begin{equation*}
\begin{aligned}
\E[Y(0)|X,Z=1] &= \E[Y(0)|X,Z=1]\Pr(Z=1|A=0)+\E[Y(0)|X,Z=1]\Pr(Z=0|A=0)\\
&=\E[Y(0)|X,Z=1]\Pr(Z=1|A=0)+\E[Y(0)|X,Z=0]\Pr(Z=0|A=0)\textit{ by Assumption~\ref{ass:mean_exchange}}\\
&=\E[Y(0)|X,A=0,Z=1]\Pr(Z=1|A=0)+\E[Y(0)|X,A=0,Z=0]\Pr(Z=0|A=0)\\
&=\E[Y|X,A=0,Z=1]\Pr(Z=1|A=0)+\E[Y|X,A=0,Z=0]\Pr(Z=0|A=0).
\end{aligned}
\end{equation*}
Using weighting identification format, $\E[Y|X,A=a,Z=z]$ for $a\in\{0,1\}$ and $z\in\{0,1\}$ can be written as
\begin{equation}
\begin{aligned}
&\E[Y|X,A=1,Z=1]=\E\left[\frac{AZY}{\Pr(A=1|Z=1,X)\Pr(Z=1|X)}\mid X\right]=\E\left[\frac{AZY}{\Pr(A=1|Z=1)\Pr(Z=1|X)}\mid X\right],\\
&\E[Y|X,A=0,Z=1]=\E\left[\frac{(1-A)ZY}{\Pr(A=0|Z=1)\Pr(Z=1|X)}\mid X\right],\\
&\E[Y|X,A=0,Z=0]=\E\left[\frac{(1-A)(1-Z)Y}{\Pr(A=0|Z=0)\Pr(Z=0|X)}\mid X\right]=\E\left[\frac{(1-A)(1-Z)Y}{\Pr(Z=0|X)}\mid X\right].
\end{aligned}
\end{equation}

Under Assumptions~\ref{ass:ignore_trt} and~\ref{ass:mean_exchange}, $\E[Y(1)-Y(0)|X,Z=0]$ equals to $\E[Y(1)-Y(0)|X,Z=1]$, because
\begin{equation*}
\begin{aligned}
&\E[Y(1)-Y(0)|X,Z=0] = \E[Y(1)|X,Z=0] - \E[Y(0)|X,Z=0]\\
=&\E[Y(1)|X,Z=1] - \E[Y(0)|X,Z=0]\Pr(Z=1|A=0) - \E[Y(0)|X,Z=0]\Pr(Z=0|A=0)\\
=&\E[Y(1)|X,Z=1] - \E[Y(0)|X,Z=1]\Pr(Z=1|A=0) - \E[Y(0)|X,Z=0]\Pr(Z=0|A=0)\\
=&\E[Y(1)-Y(0)|X,Z=1].
\end{aligned}
\end{equation*}
Hence, CATE $\tau(X)$ can be identified as
\begin{equation}
\begin{aligned}
&\tau(X)=\E[Y(1)-Y(0)|X,Z=1]\Pr(Z=1|X)+\E[Y(1)-Y(0)|X,Z=0]\Pr(Z=0|X)\\
=&\frac{\Pr(Z=1|X)\E[AZY|X]}{\Pr(A=1|Z=1)\Pr(Z=1|X)}-\frac{\Pr(Z=1|X)\E[(1-A)ZY|X]\Pr(Z=1|A=0)}{\Pr(A=0|Z=1)\Pr(Z=1|X)}\\
&-\frac{\Pr(Z=1|X)\E[(1-A)(1-Z)Y|X]\Pr(Z=0|A=0)}{\Pr(Z=0|X)}\\
&+ \frac{\Pr(Z=0|X)\E[AZY|X]}{\Pr(A=1|Z=1)\Pr(Z=1|X)}-\frac{\Pr(Z=0|X)\E[(1-A)ZY|X]\Pr(Z=1|A=0)}{\Pr(A=0|Z=1)\Pr(Z=1|X)}\\
&-\frac{\Pr(Z=0|X)\E[(1-A)(1-Z)Y|X]\Pr(Z=0|A=0)}{\Pr(Z=0|X)}\\
=&\frac{\E[AZY|X]}{\Pr(A=1|Z=1)\Pr(Z=1|X)}-\frac{\Pr(Z=1|A=0)\E[(1-A)ZY]}{\Pr(A=0|Z=1)\Pr(Z=1|X)}-\frac{\Pr(Z=0|A=0)\E[(1-A)(1-Z)Y]}{\Pr(Z=0|X)}\\
=&\frac{1}{\Pr(A=1|Z=1)}\E\left[\frac{AZY}{\Pr(Z=1|X)}\mid X\right] - \frac{\Pr(Z=1|A=0)}{\Pr(A=0|Z=1)}\E\left[\frac{(1-A)ZY}{\Pr(Z=1|X)}\mid X\right]\\
&-\Pr(Z=0|A=0)\E\left[\frac{(1-A)(1-Z)Y}{\Pr(Z=0|X)}\mid X\right]
\end{aligned}
\end{equation}

\subsection{ATI Identification}
Use tilting function $h(x)=1$, ATI can be identified as
\begin{equation}
\begin{aligned}
    \tau_\ATI &= \E[\tau(X)]\\
    &=\frac{1}{\Pr(A=1|Z=1)}\E\left[\frac{AZY}{\Pr(Z=1|X)}\right] - \frac{\Pr(Z=1|A=0)}{\Pr(A=0|Z=1)}\E\left[\frac{(1-A)ZY}{\Pr(Z=1|X)}\right]\\
&-\Pr(Z=0|A=0)\E\left[\frac{(1-A)(1-Z)Y}{\Pr(Z=0|X)}\right].
\end{aligned}
\end{equation}
Note that,
\begin{equation}
\begin{aligned}
\E\left[\frac{AZ}{\Pr(Z=1|X)}\right] &= \E\left[\E\left[\frac{AZ}{\Pr(Z=1|X)}\mid X\right]\right] = \E\left[\frac{\Pr(A=1,Z=1|X)}{\Pr(Z=1|X)}\right] \\
&= \E\left[\frac{\Pr(A=1|Z=1,X)\Pr(Z=1|X)}{\Pr(Z=1|X)}\right] = \Pr(A=1|Z=1);
\end{aligned}
\end{equation}
similarly, we have
\begin{equation}
\E\left[\frac{(1-A)Z}{\Pr(Z=1|X)}\right] = \Pr(A=0|Z=1),
\end{equation}
and
\begin{equation}
\E\left[\frac{(1-A)(1-Z)}{1-\Pr(Z=1|X)}\right] = \E\left[\E\left[\frac{(1-Z)}{1-\Pr(Z=1|X)}\mid X\right]\right] = 1.
\end{equation}
ATI identification can be written as
\begin{equation}
\begin{aligned}
    \tau_\ATI &=\frac{1}{\E\left[\frac{AZ}{\Pr(Z=1|X)}\right]}\E\left[\frac{AZY}{\Pr(Z=1|X)}\right] - \frac{\Pr(Z=1|A=0)}{\E\left[\frac{(1-A)Z}{\Pr(Z=1|X)}\right]}\E\left[\frac{(1-A)ZY}{\Pr(Z=1|X)}\right]\\
&-\frac{\Pr(Z=0|A=0)}{\E\left[\frac{(1-A)(1-Z)}{1-\Pr(Z=1|X)}\right]}\E\left[\frac{(1-A)(1-Z)Y}{\Pr(Z=0|X)}\right],
\end{aligned}
\end{equation}
which suggests an IPW estimator for ATI as
\begin{equation}
\begin{aligned}
\widehat\tau_\ATI = & \frac{\sum_{i=1}^Nw_1(X_i)A_iZ_iY_i}{\sum_{i=1}^Nw_1(X_i)A_iZ_i} - \frac{N_{10}}{N_{10}+N_2}\frac{\sum_{i=1}^N w_1(X_i)(1-A_i)Z_iY_i}{\sum_{i=1}^N w_1(X_i)(1-A_i)Z_i} \\
&- \frac{N_2}{N_{10}+N_2}\frac{\sum_{i=1}^Nw_0(X_i)(1-A_i)(1-Z_i)Y_i}{\sum_{i=1}^Nw_0(X_i)(1-A_i)(1-Z_i)},
\end{aligned}
\end{equation}
where $w_1(x)=\frac{1}{\pi(x)}$ and $w_0(x)=\frac{1}{1-\pi(x)}$.

\subsection{ATT Identification}
Use tilting function $h(x)=\pi(x)$, ATT can be identified as
\begin{equation}
\begin{aligned}
\tau_\ATT =& \frac{\E[\tau(X)\pi(X)]}{\E[\pi(X)]}\\
=&\frac{1}{\Pr(Z=1)}\E\left[\frac{\E[AZY|X]}{\Pr(A=1|Z=1)}-\frac{\Pr(Z=1|A=0)\E[(1-A)ZY|X]}{\Pr(A=0|Z=1)}\right.\\
& \left.-\Pr(Z=0|A=0)\E\left[\frac{\Pr(Z=1|X)(1-A)(1-Z)Y}{\Pr(Z=0|X)}\right]\right]\\
=&\frac{\E[AZY]}{\Pr(A=1,Z=1)}-\frac{\Pr(Z=1|A=0)\E[(1-A)ZY]}{\Pr(A=0,Z=1)}-\frac{\Pr(Z=0|A=0)}{\Pr(Z=1)}\E\left[\frac{\Pr(Z=1|X)(1-A)(1-Z)Y}{\Pr(Z=0|X)}\right]
\end{aligned}
\end{equation}
Note that, 
\begin{equation}
\begin{aligned}
    &\E\left[\frac{\Pr(Z=1|X)(1-A)(1-Z)}{1-\Pr(Z=1|X)}\right] =\E\left[\frac{\Pr(Z=1|X)(1-Z)}{1-\Pr(Z=1|X)}\right]=\E\left[\E\left[\frac{\Pr(Z=1|X)(1-Z)}{1-\Pr(Z=1|X)}\mid X\right]\right]\\
    =&\E\left[\frac{\Pr(Z=1|X)(1-\Pr(Z=1|X))}{1-\Pr(Z=1|X)}\right] = \Pr(Z=1).
\end{aligned}
\end{equation}
Then ATT can be written as
\begin{equation}
\begin{aligned}
\tau_\ATT =& \frac{\E[AZY]}{\Pr(A=1,Z=1)}-\frac{\Pr(Z=1|A=0)\E[(1-A)ZY]}{\Pr(A=0,Z=1)}\\
&-\frac{\Pr(Z=0|A=0)}{\E\left[\frac{\Pr(Z=1|X)(1-A)(1-Z)}{1-\Pr(Z=1|X)}\right] }\E\left[\frac{\Pr(Z=1|X)(1-A)(1-Z)Y}{\Pr(Z=0|X)}\right]
\end{aligned}
\end{equation}
An IPW ATT estimator can be derived
\begin{equation}
\begin{aligned}
\widehat\tau_\ATT =& \frac{\sum_{i=1}^Nw_1(X_i)A_iZ_iY_i}{\sum_{i=1}^Nw_1(X_i)A_iZ_i} - \frac{N_{10}}{N_{10}+N_2}\frac{\sum_{i=1}^Nw_1(X_i)(1-A_i)Z_iY_i}{\sum_{i=1}^Nw_1(X_i)(1-A_i)Z_i} \\
&-\frac{N_2}{N_{10}+N_2}\frac{\sum_{i=1}w_0(X_i)(1-A_i)(1-Z_i)Y_i}{\sum_{i=1}^Nw_0(X_i)(1-A_i)(1-Z_i)},
\end{aligned}
\end{equation}
where $w_1(x)=1$ and $w_0(x)=\frac{\pi(x)}{1-\pi(x)}$.

\subsection{ATO Identification}
Use tilting function $h(x)=\pi(x)(1-\pi(x))$, ATO can be identified as
\begin{equation}
\begin{aligned}
\tau_\ATO =& \frac{\E[\tau(X)\pi(X)(1-\pi(X))]}{\E[\pi(X)(1-\pi(X)]}\\
=&\frac{1}{\E[\Pr(Z=1|X)\Pr(Z=0|X)]}\E\left[\frac{\E[\Pr(Z=0|X)AZY|X]}{\Pr(A=1|Z=1)}\right.\\
&-\left.\frac{\Pr(Z=1|A=0)\E[\Pr(Z=0|X)(1-A)ZY|X]}{\Pr(A=0|Z=1)}
-\Pr(Z=0|A=0)\E[\Pr(Z=1|X)(1-A)(1-Z)Y|X]\right]\\
=&\frac{1}{\E[\Pr(Z=1|X)\Pr(Z=0|X)]}\left[\frac{\E[\Pr(Z=0|X)AZY]}{\Pr(A=1|Z=1)}\right.\\
&-\left.\frac{\Pr(Z=1|A=0)\E[\Pr(Z=0|X)(1-A)ZY]}{\Pr(A=0|Z=1)}
-\Pr(Z=0|A=0)\E[\Pr(Z=1|X)(1-A)(1-Z)Y]\right].
\end{aligned}
\end{equation}
Note that,
\begin{equation}
\begin{aligned}
\E[\Pr(Z=0|X)AZ] &= \E[\E[AZ\Pr(Z=0|X)\mid X]] = \E[\Pr(Z=0|X)\E[AZ|X]] = \E[\Pr(Z=0|X)\Pr(A=1,Z=1|X)] \\
&= \E[\Pr(Z=0|X)\Pr(A=1|Z=1,X)\Pr(Z=1|X)]\\
&= \E[\Pr(Z=0|X)\Pr(A=1|Z=1)\Pr(Z=1|X)]\\
&= \Pr(A=1|Z=1)\E[\Pr(Z=1|X)\Pr(Z=0|X)],
\end{aligned}
\end{equation}
similarly,
\begin{equation}
 \E[\Pr(Z=0|X)(1-A)Z] = \Pr(A=0|Z=1)\E[\Pr(Z=1|X)\Pr(Z=0|X)],
\end{equation}
and 
\begin{equation}
\begin{aligned}
 \E[\Pr(Z=1|X)(1-A)(1-Z)] &= \Pr(A=0|Z=0)\E[\Pr(Z=1|X)\Pr(Z=0|X)]\\
 &=\E[\Pr(Z=1|X)\Pr(Z=0|X)].
\end{aligned}
\end{equation}
Then ATO can be written as
\begin{equation}
\begin{aligned}
\tau_\ATO =& \frac{\E[\Pr(Z=0|X)AZY]}{\E[\Pr(Z=0|X)AZ]} - \frac{\Pr(Z=1|A=0)\E[\Pr(Z=0|X)(1-A)ZY]}{\E[\Pr(Z=0|X)(1-A)Z]}\\
&- \frac{\Pr(Z=0|A=0)\E[\Pr(Z=1|X)(1-A)(1-Z)Y]}{\E[\Pr(Z=1|X)(1-A)(1-Z)]}.
\end{aligned}
\end{equation}
An IPW ATO estimator can be derived
\begin{equation}
\begin{aligned}
\widehat\tau_\ATT =& \frac{\sum_{i=1}^Nw_1(X_i)A_iZ_iY_i}{\sum_{i=1}^Nw_1(X_i)A_iZ_i} - \frac{N_{10}}{N_{10}+N_2}\frac{\sum_{i=1}^Nw_1(X_i)(1-A_i)Z_iY_i}{\sum_{i=1}^Nw_1(X_i)(1-A_i)Z_i} \\
&-\frac{N_2}{N_{10}+N_2}\frac{\sum_{i=1}w_0(X_i)(1-A_i)(1-Z_i)Y_i}{\sum_{i=1}^Nw_0(X_i)(1-A_i)(1-Z_i)},
\end{aligned}
\end{equation}
where $w_1(x)=1-\pi(x)$ and $w_0(x)=\pi(x)$.

\newpage
\section{Simulation Parameters and Estimands}

\subsection{Simulation Setting}
\begin{table}[htbp]
\centering
\caption{Linear model coefficient and sample size settings.}\label{tb:coef_setting}
\begin{threeparttable}
\begin{tabular}{@{}ccccccccccccc@{}}
\toprule
\multirow{2}{*}{Setting} & \multirow{2}{*}{HTE} & \multirow{2}{*}{$N_{11}$} & \multirow{2}{*}{$N_{10}$} & \multirow{2}{*}{$N_2$} & \multirow{2}{*}{$N_{10}:N_2$} & \multirow{2}{*}{$\lambda=\frac{N_1}{N_1+N_2}$} & \multicolumn{6}{c}{Outcome Model Parameters}  \\ \cline{8-13} 
&&&&&&& $\beta_0$ & $\beta_1$ & $\beta_2$ & $\beta_{\text{trt}}$ & $\phi_1$ & $\phi_2$               \\ \hline
\multicolumn{13}{c}{Case 1: RCT ($N_1=200$) has a 1:1 treatment-control allocation ratio}                \\ \hline
1  &              & 100     & 100 & 100   & 1:1  & 2/3 & 0    & 1    & 1    & 0    & 0.00 & 0.00 \\
2  & $\checkmark$ & 100     & 100 & 100   & 1:1  & 2/3 & 0    & 1    & 1    & 0    & 0.25 & 0.25 \\
3  & $\checkmark$ & 100     & 100 & 100   & 1:1  & 2/3 & 0    & 1    & 1    & 0    & 0.50 & 0.50 \\
4  &              & 100     & 100 & 300   & 1:3  & 2/5 & 0    & 1    & 1    & 0    & 0.00 & 0.00 \\
5  & $\checkmark$ & 100     & 100 & 300   & 1:3  & 2/5 & 0    & 1    & 1    & 0    & 0.25 & 0.25 \\
6  & $\checkmark$ & 100     & 100 & 300   & 1:3  & 2/5 & 0    & 1    & 1    & 0    & 0.50 & 0.50 \\ 
7  &              & 100     & 100 & 1000  & 1:10 & 1/6 & 0    & 1    & 1    & 0    & 0.00 & 0.00 \\
8  & $\checkmark$ & 100     & 100 & 1000  & 1:10 & 1/6 & 0    & 1    & 1    & 0    & 0.25 & 0.25 \\
9  & $\checkmark$ & 100     & 100 & 1000  & 1:10 & 1/6 & 0    & 1    & 1    & 0    & 0.50 & 0.50 \\ \hline
\multicolumn{13}{c}{Case 2: RCT ($N_1=200$) has a 3:1 treatment-control allocation ratio}                \\ \hline
10  &              & 150     & 50 & 50    & 1:1  & 4/5 & 0    & 1    & 1    & 0    & 0.00 & 0.00 \\
11  & $\checkmark$ & 150     & 50 & 50    & 1:1  & 4/5 & 0    & 1    & 1    & 0    & 0.25 & 0.25 \\
12  & $\checkmark$ & 150     & 50 & 50    & 1:1  & 4/5 & 0    & 1    & 1    & 0    & 0.50 & 0.50 \\
13  &              & 150     & 50 & 150   & 1:3  & 4/7 & 0    & 1    & 1    & 0    & 0.00 & 0.00 \\
14  & $\checkmark$ & 150     & 50 & 150   & 1:3  & 4/7 & 0    & 1    & 1    & 0    & 0.25 & 0.25 \\
15  & $\checkmark$ & 150     & 50 & 150   & 1:3  & 4/7 & 0    & 1    & 1    & 0    & 0.50 & 0.50 \\ 
16  &               & 150     & 50 & 500   & 1:10 & 2/7 & 0    & 1    & 1    & 0    & 0.00 & 0.00 \\
17  & $\checkmark$  & 150     & 50 & 500   & 1:10 & 2/7 & 0    & 1    & 1    & 0    & 0.25 & 0.25 \\
18  & $\checkmark$  & 150     & 50 & 500   & 1:10 & 2/7 & 0    & 1    & 1    & 0    & 0.50 & 0.50 \\ \hline
\end{tabular}
\end{threeparttable}
\end{table}

\subsection{Estimand Calculation Procedure in Simulation}\label{app:compute_estimand}

True values of causal estimand are computed using Monte Carlo integration as the following steps:
\begin{enumerate}
    \item Generate covariates $X_1$ and $X_2$ independently for the RCT population with size $n_1=10^6$ and for the EC population with size $n_2=\frac{\lambda}{1-\lambda}\times10^6$. The size of the EC population involves $\lambda$ to reflect the mixture proportion of the integrated population.
    \item Compute the joint densities $f_1(x)$ and $f_2(x)$ as the product of the marginal densities for $X_1$ and $X_2$, assuming independence. That is, $f_j(x)=f_{j,X_1}(x_1)f_{j,X_2}(x_2)$ with $j=1$ for RCT and $j=2$ for EC, $f_{j,X_1}(x_1)$ and $f_{j,X_2}(x_2)$ stand for $X_1$ and $X_2$ density functions for population $j$, respectively.
    \item Compute true PS $\pi(x)$ using Eq~\eqref{eq:pi(x)}.
    \item Compute tilting functions $h(x)$ for ATI, ATT and ATO using formula in Table~\ref{tb:weights}.
    \item Compute $\tau(x_i)$ as $\tau(x_i)= \beta_{\text{trt}}+\phi_1x_{1i}+\phi_2 x_{2i}$ for each generated unit $i\in\{1,\ldots,n_1+n_2\}$.
    \item Approximate true estimand values as $\tau_{k}\approx \frac{\sum_{i=1}^{n_1+n_2}h(x_i)\tau(x_i)}{\sum_{i=1}^{n_1+n_2}h(x_i)}$, with $k\in\{\ATI,\ATT,\ATO\}$.
\end{enumerate}

\newpage
\section{Additional Simulation Results}\label{app:add_sim_results}

\begin{figure}[htbp]
    \centering
    \includegraphics[width=\linewidth]{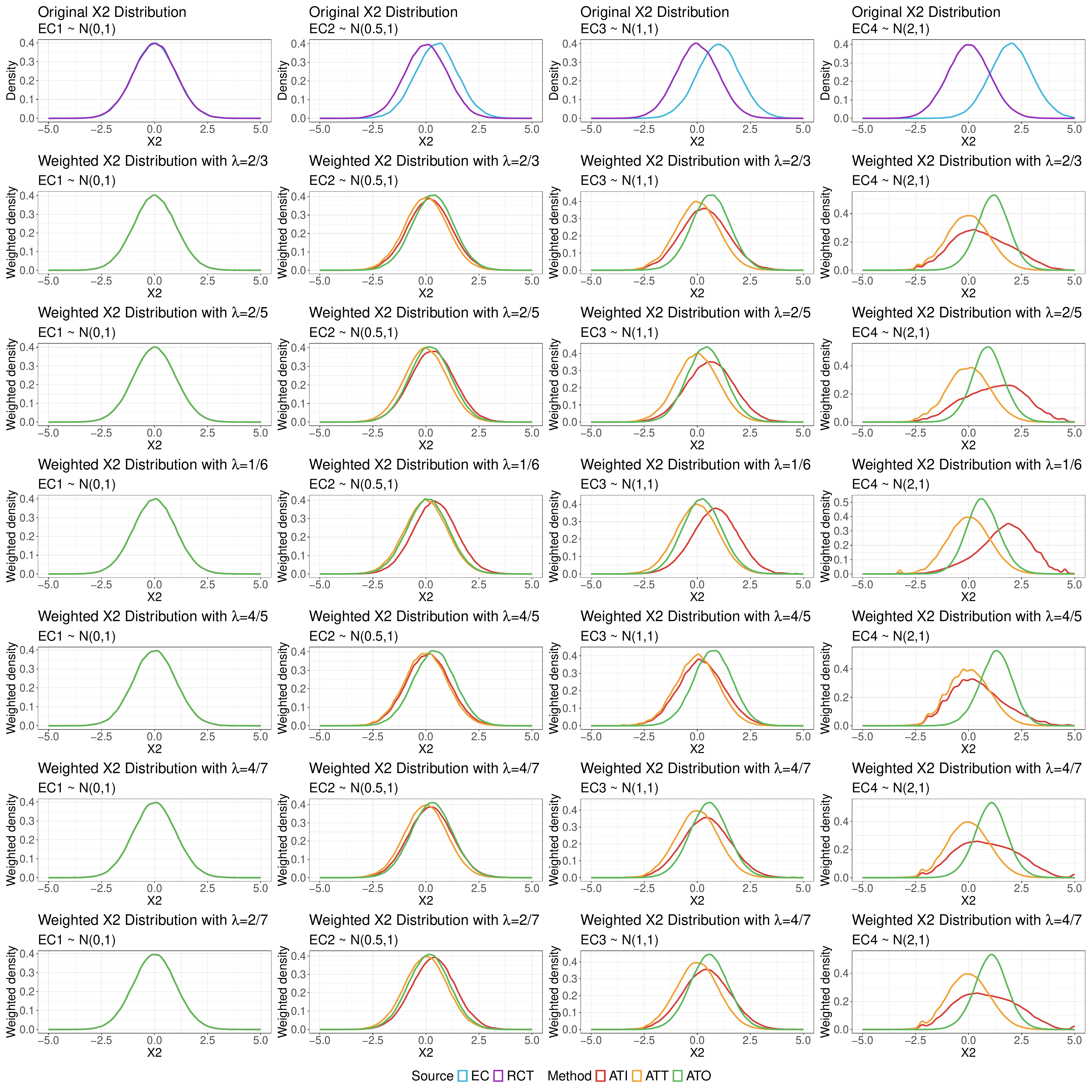}
    \caption{Density plot of original $X_2$ and weighted $X_2$ of RCT and EC super-populations with 6 different mixture proportion $\lambda$ as mentioned in Table~\ref{tb:coef_setting}. Four types of ECs are presented in this Figure: EC1 with $X_2\sim N(0,1)$, EC2 with $X_2\sim N(0.5,1)$, EC3 with $X_2\sim N(1,1)$ and EC4 with $X_2\sim N(2,1)$. Weighted distributions for $X_2$ are derived using $w_1(x)$ and $w_0(x)$ in Table~\ref{tb:weights}.}
    \label{fig:x2_1}
\end{figure}

\begin{figure}[htbp]
    \centering
    \includegraphics[width=\linewidth]{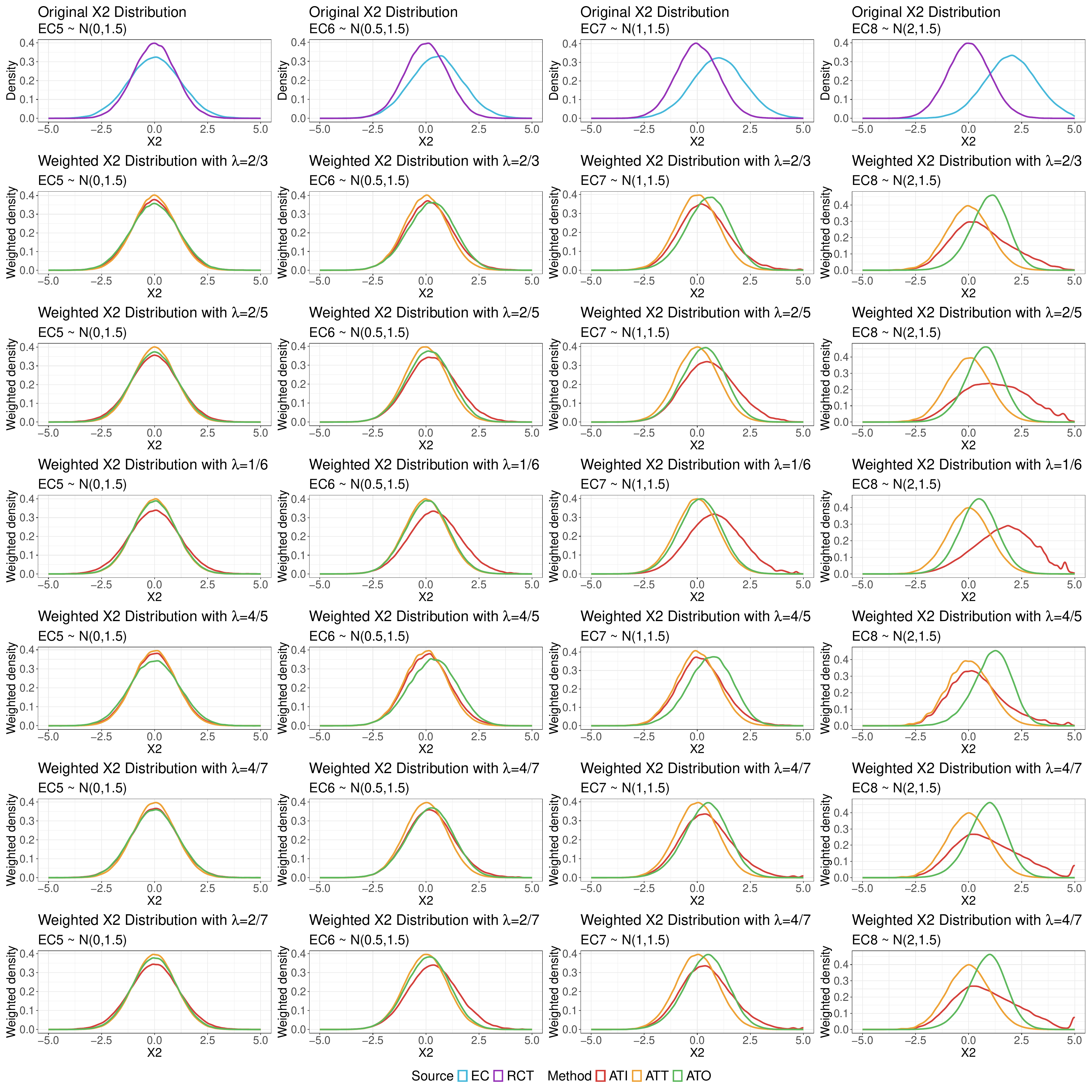}
    \caption{Density plot of original $X_2$ and weighted $X_2$ of RCT and EC super-populations with 6 different mixture proportion $\lambda$ as mentioned in Table~\ref{tb:coef_setting}. Four types of ECs are presented in this Figure: EC5 with $X_2\sim N(0,1.5)$, EC6 with $X_2\sim N(0.5,1.5)$, EC7 with $X_2\sim N(1,1.5)$ and EC8 with $X_2\sim N(2,1.5)$. Weighted distributions for $X_2$ are derived using $w_1(x)$ and $w_0(x)$ in Table~\ref{tb:weights}.}
    \label{fig:x2_2}
\end{figure}

\begin{figure}[htbp]
    \centering
    \includegraphics[width=\linewidth]{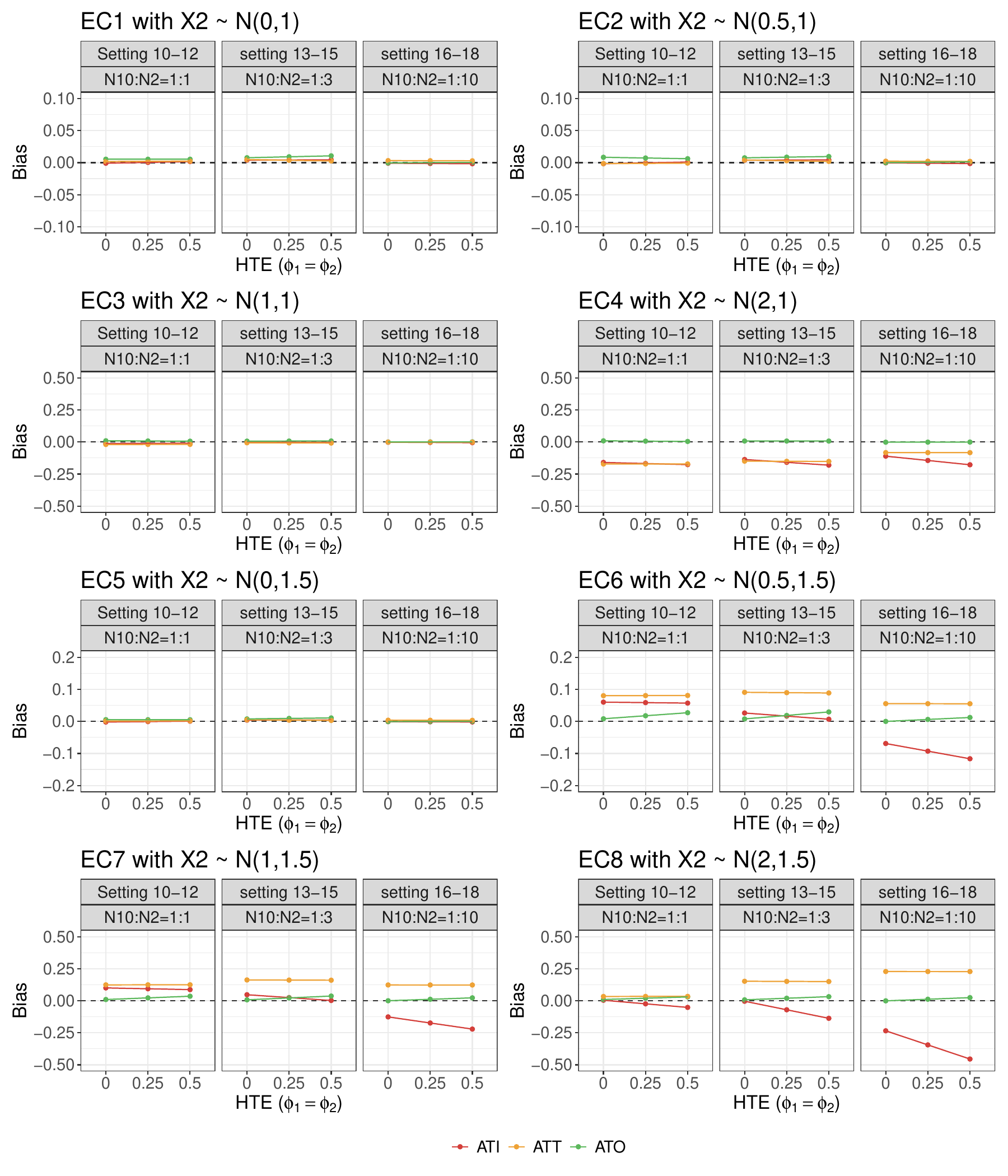}
    \caption{Bias of proposed estimators in estimating ATI, ATT and ATO under 3:1 treatment-control allocation ratio of RCT (setting 10 to 18 in Table~\ref{tb:coef_setting}).}
    \label{fig:bias2}
\end{figure}

\begin{figure}[htbp]
    \centering
    \includegraphics[width=\linewidth]{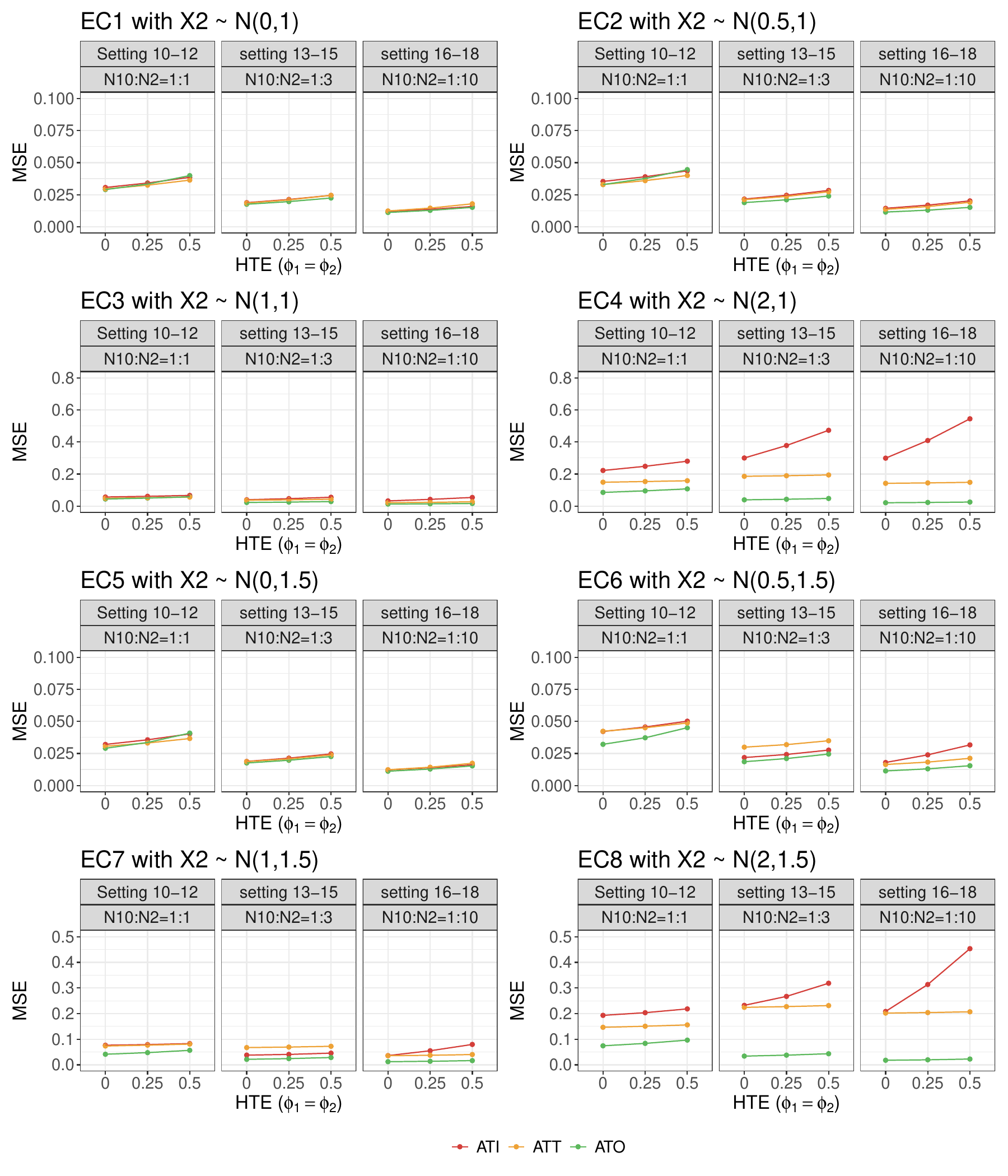}
    \caption{MSE of proposed estimators in estimating ATI, ATT and ATO under 3:1 treatment-control allocation ratio of RCT (setting 10 to 18 in Table~\ref{tb:coef_setting}).}
    \label{fig:mse2}
\end{figure}

\newpage

{\small
\centering
\begin{longtable}[htbp]{lrrrlrrrlrrrlrrr}
\caption{Detailed simulation results for average estimands.}\label{tb:app_sim_estimand}\\
\toprule
   &        \multicolumn{3}{c}{Estimand} &
   &        \multicolumn{3}{c}{Estimand} &
   &        \multicolumn{3}{c}{Estimand} &
   &        \multicolumn{3}{c}{Estimand} \\
\cmidrule(lr){2-4}\cmidrule(lr){6-8}\cmidrule(lr){10-12}\cmidrule(lr){14-16}
EC & ATI & ATT & ATO &
EC & ATI & ATT & ATO &
EC & ATI & ATT & ATO &
EC & ATI & ATT & ATO \\
\midrule
\endfirsthead
\toprule
   &        \multicolumn{3}{c}{Estimand} &
   &        \multicolumn{3}{c}{Estimand} &
   &        \multicolumn{3}{c}{Estimand} &
   &        \multicolumn{3}{c}{Estimand} \\
\cmidrule(lr){2-4}\cmidrule(lr){6-8}\cmidrule(lr){10-12}\cmidrule(lr){14-16}
EC & ATI & ATT & ATO &
EC & ATI & ATT & ATO &
EC & ATI & ATT & ATO &
EC & ATI & ATT & ATO\\
\midrule
\endhead

\midrule
\multicolumn{16}{r}{{Continued on next page}} \\
\midrule
\endfoot

\bottomrule
\endlastfoot

\multicolumn{16}{l}{Setting 1: $\phi_1=\phi_2=0$ and $\lambda = 2/3$}   \\ \hline
EC1 & 0.00 & 0.00 & 0.00 &
EC2 & 0.00 & 0.00 & 0.00 &
EC3 & 0.00 & 0.00 & 0.00 &
EC4 & 0.00 & 0.00 & 0.00 \\
EC5 & 0.00 & 0.00 & 0.00 &
EC6 & 0.00 & 0.00 & 0.00 &
EC7 & 0.00 & 0.00 & 0.00 &
EC8 & 0.00 & 0.00 & 0.00 \\\hline
\multicolumn{16}{l}{Setting 2: $\phi_1=\phi_2=0.25$ and $\lambda = 2/3$}   \\ \hline
EC1 & 0.12 & 0.12 & 0.12 &
EC2 & 0.17 & 0.12 & 0.21 &
EC3 & 0.21 & 0.12 & 0.28 &
EC4 & 0.29 & 0.12 & 0.41 \\
EC5 & 0.12 & 0.12 & 0.12 &
EC6 & 0.17 & 0.12 & 0.18 &
EC7 & 0.21 & 0.12 & 0.25 &
EC8 & 0.29 & 0.12 & 0.36 \\\hline
\multicolumn{16}{l}{Setting 3: $\phi_1=\phi_2=0.5$ and $\lambda = 2/3$}   \\ \hline
EC1 & 0.25 & 0.25 & 0.25 &
EC2 & 0.33 & 0.25 & 0.41 &
EC3 & 0.42 & 0.25 & 0.56 &
EC4 & 0.58 & 0.25 & 0.82 \\
EC5 & 0.25 & 0.25 & 0.25 &
EC6 & 0.33 & 0.25 & 0.37 &
EC7 & 0.42 & 0.25 & 0.49 &
EC8 & 0.58 & 0.25 & 0.73 \\\hline
\multicolumn{16}{l}{Setting 4: $\phi_1=\phi_2=0$ and $\lambda = 2/5$}   \\ \hline
EC1 & 0.00 & 0.00 & 0.00 &
EC2 & 0.00 & 0.00 & 0.00 &
EC3 & 0.00 & 0.00 & 0.00 &
EC4 & 0.00 & 0.00 & 0.00 \\
EC5 & 0.00 & 0.00 & 0.00 &
EC6 & 0.00 & 0.00 & 0.00 &
EC7 & 0.00 & 0.00 & 0.00 &
EC8 & 0.00 & 0.00 & 0.00 \\\hline
\multicolumn{16}{l}{Setting 5: $\phi_1=\phi_2=0.25$ and $\lambda = 2/5$}   \\ \hline
EC1 & 0.12 & 0.12 & 0.12 &
EC2 & 0.20 & 0.12 & 0.18 &
EC3 & 0.27 & 0.12 & 0.23 &
EC4 & 0.42 & 0.12 & 0.35 \\
EC5 & 0.12 & 0.12 & 0.12 &
EC6 & 0.20 & 0.12 & 0.16 &
EC7 & 0.27 & 0.12 & 0.20 &
EC8 & 0.42 & 0.12 & 0.30 \\\hline
\multicolumn{16}{l}{Setting 6: $\phi_1=\phi_2=0.5$ and $\lambda = 2/5$}   \\ \hline
EC1 & 0.25 & 0.25 & 0.25 &
EC2 & 0.40 & 0.25 & 0.35 &
EC3 & 0.55 & 0.25 & 0.46 &
EC4 & 0.85 & 0.25 & 0.71 \\
EC5 & 0.25 & 0.25 & 0.25 &
EC6 & 0.40 & 0.25 & 0.31 &
EC7 & 0.55 & 0.25 & 0.40 &
EC8 & 0.85 & 0.25 & 0.61 \\\hline
\multicolumn{16}{l}{Setting 7: $\phi_1=\phi_2=0$ and $\lambda = 1/6$}   \\ \hline
EC1 & 0.00 & 0.00 & 0.00 &
EC2 & 0.00 & 0.00 & 0.00 &
EC3 & 0.00 & 0.00 & 0.00 &
EC4 & 0.00 & 0.00 & 0.00 \\
EC5 & 0.00 & 0.00 & 0.00 &
EC6 & 0.00 & 0.00 & 0.00 &
EC7 & 0.00 & 0.00 & 0.00 &
EC8 & 0.00 & 0.00 & 0.00 \\\hline
\multicolumn{16}{l}{Setting 8: $\phi_1=\phi_2=0.25$ and $\lambda = 1/6$}   \\ \hline
EC1 & 0.12 & 0.12 & 0.12 &
EC2 & 0.23 & 0.12 & 0.15 &
EC3 & 0.33 & 0.12 & 0.19 &
EC4 & 0.54 & 0.12 & 0.29 \\
EC5 & 0.12 & 0.12 & 0.12 &
EC6 & 0.23 & 0.12 & 0.14 &
EC7 & 0.33 & 0.12 & 0.16 &
EC8 & 0.54 & 0.12 & 0.24 \\\hline
\multicolumn{16}{l}{Setting 9: $\phi_1=\phi_2=0.5$ and $\lambda = 1/6$}   \\ \hline
EC1 & 0.25 & 0.25 & 0.25 &
EC2 & 0.46 & 0.25 & 0.30 &
EC3 & 0.67 & 0.25 & 0.37 &
EC4 & 1.08 & 0.25 & 0.59 \\
EC5 & 0.25 & 0.25 & 0.25 &
EC6 & 0.46 & 0.25 & 0.28 &
EC7 & 0.67 & 0.25 & 0.32 &
EC8 & 1.08 & 0.25 & 0.48 \\\hline
\multicolumn{16}{l}{Setting 10: $\phi_1=\phi_2=0$ and $\lambda = 4/5$}   \\ \hline
EC1 & 0.00 & 0.00 & 0.00 &
EC2 & 0.00 & 0.00 & 0.00 &
EC3 & 0.00 & 0.00 & 0.00 &
EC4 & 0.00 & 0.00 & 0.00 \\
EC5 & 0.00 & 0.00 & 0.00 &
EC6 & 0.00 & 0.00 & 0.00 &
EC7 & 0.00 & 0.00 & 0.00 &
EC8 & 0.00 & 0.00 & 0.00 \\\hline
\multicolumn{16}{l}{Setting 11: $\phi_1=\phi_2=0.25$ and $\lambda = 4/5$}   \\ \hline
EC1 & 0.12 & 0.12 & 0.12 &
EC2 & 0.15 & 0.12 & 0.22 &
EC3 & 0.17 & 0.12 & 0.31 &
EC4 & 0.22 & 0.12 & 0.44 \\
EC5 & 0.12 & 0.12 & 0.12 &
EC6 & 0.15 & 0.12 & 0.20 &
EC7 & 0.17 & 0.12 & 0.27 &
EC8 & 0.22 & 0.12 & 0.40 \\\hline
\multicolumn{16}{l}{Setting 12: $\phi_1=\phi_2=0.5$ and $\lambda = 4/5$}   \\ \hline
EC1 & 0.25 & 0.25 & 0.25 &
EC2 & 0.30 & 0.25 & 0.44 &
EC3 & 0.35 & 0.25 & 0.61 &
EC4 & 0.45 & 0.25 & 0.89 \\
EC5 & 0.25 & 0.25 & 0.25 &
EC6 & 0.30 & 0.25 & 0.40 &
EC7 & 0.35 & 0.25 & 0.55 &
EC8 & 0.45 & 0.25 & 0.80 \\\hline
\multicolumn{16}{l}{Setting 13: $\phi_1=\phi_2=0$ and $\lambda = 4/7$}   \\ \hline
EC1 & 0.00 & 0.00 & 0.00 &
EC2 & 0.00 & 0.00 & 0.00 &
EC3 & 0.00 & 0.00 & 0.00 &
EC4 & 0.00 & 0.00 & 0.00 \\
EC5 & 0.00 & 0.00 & 0.00 &
EC6 & 0.00 & 0.00 & 0.00 &
EC7 & 0.00 & 0.00 & 0.00 &
EC8 & 0.00 & 0.00 & 0.00 \\\hline
\multicolumn{16}{l}{Setting 14: $\phi_1=\phi_2=0.25$ and $\lambda = 4/7$}   \\ \hline
EC1 & 0.12 & 0.12 & 0.12 &
EC2 & 0.18 & 0.12 & 0.20 &
EC3 & 0.23 & 0.12 & 0.26 &
EC4 & 0.34 & 0.12 & 0.39 \\
EC5 & 0.12 & 0.12 & 0.12 &
EC6 & 0.18 & 0.12 & 0.17 &
EC7 & 0.23 & 0.12 & 0.23 &
EC8 & 0.34 & 0.12 & 0.34 \\\hline
\multicolumn{16}{l}{Setting 15: $\phi_1=\phi_2=0.5$ and $\lambda = 4/7$}   \\ \hline
EC1 & 0.25 & 0.25 & 0.25 &
EC2 & 0.36 & 0.25 & 0.39 &
EC3 & 0.46 & 0.25 & 0.53 &
EC4 & 0.68 & 0.25 & 0.78 \\
EC5 & 0.25 & 0.25 & 0.25 &
EC6 & 0.36 & 0.25 & 0.35 &
EC7 & 0.46 & 0.25 & 0.45 &
EC8 & 0.68 & 0.25 & 0.68 \\\hline
\multicolumn{16}{l}{Setting 16: $\phi_1=\phi_2=0$ and $\lambda = 2/7$}   \\ \hline
EC1 & 0.00 & 0.00 & 0.00 &
EC2 & 0.00 & 0.00 & 0.00 &
EC3 & 0.00 & 0.00 & 0.00 &
EC4 & 0.00 & 0.00 & 0.00 \\
EC5 & 0.00 & 0.00 & 0.00 &
EC6 & 0.00 & 0.00 & 0.00 &
EC7 & 0.00 & 0.00 & 0.00 &
EC8 & 0.00 & 0.00 & 0.00 \\\hline
\multicolumn{16}{l}{Setting 17: $\phi_1=\phi_2=0.25$ and $\lambda = 2/7$}   \\ \hline
EC1 & 0.13 & 0.12 & 0.13 &
EC2 & 0.21 & 0.12 & 0.16 &
EC3 & 0.30 & 0.12 & 0.21 &
EC4 & 0.48 & 0.12 & 0.33 \\
EC5 & 0.13 & 0.12 & 0.13 &
EC6 & 0.21 & 0.12 & 0.15 &
EC7 & 0.30 & 0.12 & 0.18 &
EC8 & 0.48 & 0.12 & 0.28 \\\hline
\multicolumn{16}{l}{Setting 18: $\phi_1=\phi_2=0.5$ and $\lambda = 2/7$}   \\ \hline
EC1 & 0.25 & 0.25 & 0.25 &
EC2 & 0.43 & 0.25 & 0.33 &
EC3 & 0.61 & 0.25 & 0.42 &
EC4 & 0.96 & 0.25 & 0.66 \\
EC5 & 0.25 & 0.25 & 0.25 &
EC6 & 0.43 & 0.25 & 0.29 &
EC7 & 0.61 & 0.25 & 0.36 &
EC8 & 0.96 & 0.25 & 0.55 \\
\bottomrule
\end{longtable}
}

\newpage

{
\small
\centering
\begin{longtable}[htbp]{lrrrlrrrlrrrlrrr}
\caption{Detailed simulation results for average bias.}\label{tb:app_sim_bias}\\
\toprule
   &        \multicolumn{3}{c}{Bias} &
   &        \multicolumn{3}{c}{Bias} &
   &        \multicolumn{3}{c}{Bias} &
   &        \multicolumn{3}{c}{Bias} \\
\cmidrule(lr){2-4}\cmidrule(lr){6-8}\cmidrule(lr){10-12}\cmidrule(lr){14-16}
EC & ATI & ATT & ATO &
EC & ATI & ATT & ATO &
EC & ATI & ATT & ATO &
EC & ATI & ATT & ATO \\
\midrule
\endfirsthead
\toprule
   &        \multicolumn{3}{c}{Bias} &
   &        \multicolumn{3}{c}{Bias} &
   &        \multicolumn{3}{c}{Bias} &
   &        \multicolumn{3}{c}{Bias} \\
\cmidrule(lr){2-4}\cmidrule(lr){6-8}\cmidrule(lr){10-12}\cmidrule(lr){14-16}
EC & ATI & ATT & ATO &
EC & ATI & ATT & ATO &
EC & ATI & ATT & ATO &
EC & ATI & ATT & ATO \\
\midrule
\endhead

\midrule
\multicolumn{16}{r}{{Continued on next page}} \\
\midrule
\endfoot

\bottomrule
\endlastfoot

\multicolumn{16}{l}{Setting 1: $\phi_1=\phi_2=0$ and $\lambda = 2/3$}   \\ \hline
EC1 & 0.00 & 0.00 & 0.01 & 
EC2 & 0.00 & 0.00 & 0.01 & 
EC3 & -0.01 & 0.00 & 0.01 & 
EC4 & -0.13 & -0.10 & 0.01 \\ 
EC5 & 0.00 & 0.00 & 0.01 & 
EC6 & 0.03 & 0.07 & 0.01 & 
EC7 & 0.06 & 0.13 & 0.01 & 
EC8 & 0.01 & 0.11 & 0.01 \\ \hline
\multicolumn{16}{l}{Setting 2: $\phi_1=\phi_2=0.25$ and $\lambda = 2/3$}   \\ \hline
EC1 & -0.01 & 0.00 & 0.01 & 
EC2 & -0.01 & 0.00 & 0.01 & 
EC3 & -0.01 & 0.00 & 0.01 & 
EC4 & -0.15 & -0.10 & 0.01 \\ 
EC5 & -0.01 & 0.01 & 0.01 & 
EC6 & 0.02 & 0.07 & 0.02 & 
EC7 & 0.04 & 0.13 & 0.02 & 
EC8 & -0.04 & 0.11 & 0.02 \\ \hline
\multicolumn{16}{l}{Setting 3: $\phi_1=\phi_2=0.5$ and $\lambda = 2/3$}   \\ \hline
EC1 & -0.01 & 0.00 & 0.01 & 
EC2 & -0.01 & 0.00 & 0.01 & 
EC3 & -0.01 & 0.00 & 0.01 & 
EC4 & -0.17 & -0.10 & 0.01 \\ 
EC5 & -0.01 & 0.01 & 0.01 & 
EC6 & 0.01 & 0.07 & 0.03 & 
EC7 & 0.02 & 0.13 & 0.04 & 
EC8 & -0.10 & 0.11 & 0.04 \\ \hline
\multicolumn{16}{l}{Setting 4: $\phi_1=\phi_2=0$ and $\lambda = 2/5$}   \\ \hline
EC1 & 0.00 & 0.00 & 0.01 & 
EC2 & 0.00 & 0.00 & 0.01 & 
EC3 & 0.00 & -0.01 & 0.00 & 
EC4 & -0.14 & -0.09 & 0.00 \\ 
EC5 & 0.00 & 0.00 & 0.01 & 
EC6 & -0.03 & 0.06 & 0.01 & 
EC7 & -0.04 & 0.12 & 0.00 & 
EC8 & -0.11 & 0.19 & 0.00 \\ \hline
\multicolumn{16}{l}{Setting 5: $\phi_1=\phi_2=0.25$ and $\lambda = 2/5$}   \\ \hline
EC1 & 0.00 & 0.00 & 0.01 & 
EC2 & 0.00 & 0.00 & 0.01 & 
EC3 & -0.01 & -0.01 & 0.00 & 
EC4 & -0.18 & -0.09 & 0.00 \\ 
EC5 & 0.00 & 0.00 & 0.01 & 
EC6 & -0.04 & 0.06 & 0.01 & 
EC7 & -0.08 & 0.12 & 0.02 & 
EC8 & -0.21 & 0.19 & 0.01 \\ \hline
\multicolumn{16}{l}{Setting 6: $\phi_1=\phi_2=0.5$ and $\lambda = 2/5$}   \\ \hline
EC1 & 0.00 & 0.00 & 0.01 & 
EC2 & 0.00 & 0.00 & 0.01 & 
EC3 & -0.01 & -0.01 & 0.00 & 
EC4 & -0.22 & -0.10 & 0.00 \\ 
EC5 & 0.00 & 0.00 & 0.01 & 
EC6 & -0.06 & 0.06 & 0.02 & 
EC7 & -0.12 & 0.12 & 0.03 & 
EC8 & -0.31 & 0.19 & 0.03 \\ \hline
\multicolumn{16}{l}{Setting 7: $\phi_1=\phi_2=0$ and $\lambda = 1/6$}   \\ \hline
EC1 & 0.00 & 0.00 & -0.01 & 
EC2 & 0.00 & 0.00 & -0.01 & 
EC3 & -0.01 & 0.00 & -0.01 & 
EC4 & -0.19 & -0.05 & -0.02 \\ 
EC5 & 0.00 & 0.00 & -0.01 & 
EC6 & -0.10 & 0.03 & -0.01 & 
EC7 & -0.20 & 0.08 & -0.01 & 
EC8 & -0.40 & 0.21 & -0.01 \\ \hline
\multicolumn{16}{l}{Setting 8: $\phi_1=\phi_2=0.25$ and $\lambda = 1/6$}   \\ \hline
EC1 & 0.00 & 0.00 & -0.01 & 
EC2 & 0.00 & 0.00 & -0.01 & 
EC3 & -0.01 & 0.00 & -0.01 & 
EC4 & -0.24 & -0.05 & -0.02 \\ 
EC5 & 0.00 & 0.00 & -0.01 & 
EC6 & -0.14 & 0.03 & -0.01 & 
EC7 & -0.26 & 0.08 & 0.00 & 
EC8 & -0.53 & 0.21 & 0.00 \\ \hline
\multicolumn{16}{l}{Setting 9: $\phi_1=\phi_2=0.5$ and $\lambda = 1/6$}   \\ \hline
EC1 & 0.00 & 0.00 & -0.01 & 
EC2 & 0.00 & 0.00 & -0.01 & 
EC3 & -0.01 & 0.00 & -0.01 & 
EC4 & -0.29 & -0.05 & -0.02 \\ 
EC5 & 0.00 & 0.00 & -0.01 & 
EC6 & -0.17 & 0.03 & 0.00 & 
EC7 & -0.32 & 0.08 & 0.00 & 
EC8 & -0.67 & 0.21 & 0.01 \\ \hline
\multicolumn{16}{l}{Setting 10: $\phi_1=\phi_2=0$ and $\lambda = 4/5$}   \\ \hline
EC1 & 0.00 & 0.00 & 0.01 & 
EC2 & 0.00 & 0.00 & 0.01 & 
EC3 & -0.01 & -0.02 & 0.01 & 
EC4 & -0.16 & -0.17 & 0.01 \\ 
EC5 & 0.00 & 0.00 & 0.01 & 
EC6 & 0.06 & 0.08 & 0.01 & 
EC7 & 0.10 & 0.12 & 0.01 & 
EC8 & 0.00 & 0.03 & 0.01 \\ \hline
\multicolumn{16}{l}{Setting 11: $\phi_1=\phi_2=0.25$ and $\lambda = 4/5$}   \\ \hline
EC1 & 0.00 & 0.00 & 0.01 & 
EC2 & 0.00 & 0.00 & 0.01 & 
EC3 & -0.01 & -0.02 & 0.01 & 
EC4 & -0.17 & -0.17 & 0.01 \\ 
EC5 & 0.00 & 0.00 & 0.01 & 
EC6 & 0.06 & 0.08 & 0.02 & 
EC7 & 0.09 & 0.12 & 0.02 & 
EC8 & -0.02 & 0.03 & 0.02 \\ \hline
\multicolumn{16}{l}{Setting 12: $\phi_1=\phi_2=0.5$ and $\lambda = 4/5$}   \\ \hline
EC1 & 0.00 & 0.00 & 0.01 & 
EC2 & 0.00 & 0.00 & 0.01 & 
EC3 & -0.01 & -0.02 & 0.01 & 
EC4 & -0.18 & -0.17 & 0.00 \\ 
EC5 & 0.00 & 0.00 & 0.01 & 
EC6 & 0.06 & 0.08 & 0.03 & 
EC7 & 0.09 & 0.12 & 0.04 & 
EC8 & -0.05 & 0.03 & 0.03 \\ \hline
\multicolumn{16}{l}{Setting 13: $\phi_1=\phi_2=0$ and $\lambda = 4/7$}   \\ \hline
EC1 & 0.00 & 0.00 & 0.01 & 
EC2 & 0.00 & 0.00 & 0.01 & 
EC3 & 0.00 & -0.01 & 0.01 & 
EC4 & -0.14 & -0.15 & 0.01 \\ 
EC5 & 0.00 & 0.01 & 0.01 & 
EC6 & 0.03 & 0.09 & 0.01 & 
EC7 & 0.05 & 0.16 & 0.01 & 
EC8 & 0.00 & 0.15 & 0.01 \\ \hline
\multicolumn{16}{l}{Setting 14: $\phi_1=\phi_2=0.25$ and $\lambda = 4/7$}   \\ \hline
EC1 & 0.00 & 0.00 & 0.01 & 
EC2 & 0.00 & 0.00 & 0.01 & 
EC3 & 0.00 & -0.01 & 0.01 & 
EC4 & -0.16 & -0.15 & 0.01 \\ 
EC5 & 0.00 & 0.00 & 0.01 & 
EC6 & 0.02 & 0.09 & 0.02 & 
EC7 & 0.02 & 0.16 & 0.02 & 
EC8 & -0.07 & 0.15 & 0.02 \\ \hline
\multicolumn{16}{l}{Setting 15: $\phi_1=\phi_2=0.5$ and $\lambda = 4/7$}   \\ \hline
EC1 & 0.00 & 0.00 & 0.01 & 
EC2 & 0.00 & 0.00 & 0.01 & 
EC3 & 0.00 & -0.01 & 0.01 & 
EC4 & -0.18 & -0.15 & 0.01 \\ 
EC5 & 0.00 & 0.00 & 0.01 & 
EC6 & 0.01 & 0.09 & 0.03 & 
EC7 & 0.00 & 0.16 & 0.04 & 
EC8 & -0.14 & 0.15 & 0.03 \\ \hline
\multicolumn{16}{l}{Setting 16: $\phi_1=\phi_2=0$ and $\lambda = 2/7$}   \\ \hline
EC1 & 0.00 & 0.00 & 0.00 & 
EC2 & 0.00 & 0.00 & 0.00 & 
EC3 & 0.00 & 0.00 & 0.00 & 
EC4 & -0.11 & -0.08 & 0.00 \\ 
EC5 & 0.00 & 0.00 & 0.00 & 
EC6 & -0.07 & 0.06 & 0.00 & 
EC7 & -0.13 & 0.12 & 0.00 & 
EC8 & -0.23 & 0.23 & 0.00 \\ \hline
\multicolumn{16}{l}{Setting 17: $\phi_1=\phi_2=0.25$ and $\lambda = 2/7$}   \\ \hline
EC1 & 0.00 & 0.00 & 0.00 & 
EC2 & 0.00 & 0.00 & 0.00 & 
EC3 & 0.00 & 0.00 & 0.00 & 
EC4 & -0.14 & -0.08 & 0.00 \\ 
EC5 & 0.00 & 0.00 & 0.00 & 
EC6 & -0.09 & 0.06 & 0.01 & 
EC7 & -0.17 & 0.12 & 0.01 & 
EC8 & -0.34 & 0.23 & 0.01 \\ \hline
\multicolumn{16}{l}{Setting 18: $\phi_1=\phi_2=0.5$ and $\lambda = 2/7$}   \\ \hline
EC1 & 0.00 & 0.00 & 0.00 & 
EC2 & 0.00 & 0.00 & 0.00 & 
EC3 & -0.01 & 0.00 & 0.00 & 
EC4 & -0.18 & -0.08 & 0.00 \\ 
EC5 & 0.00 & 0.00 & 0.00 & 
EC6 & -0.12 & 0.06 & 0.01 & 
EC7 & -0.22 & 0.12 & 0.02 & 
EC8 & -0.45 & 0.23 & 0.02 \\
\bottomrule
\end{longtable}
}

\end{appendices}

\newpage
\bibliographystyle{apalike}
\bibliography{sample}

\end{document}